\documentclass[aip,pof,longbibliography,twocolumn,reprint]{revtex4-1}

\usepackage{amssymb,amsmath}

\usepackage{graphicx}

\usepackage{color}
\usepackage{xcolor}
\newcommand{\vicente}[1]{{ #1}}

\newcommand\beq{\begin{equation}}
\newcommand\eeq{\end{equation}}
\newcommand\beqa{\begin{eqnarray}}
\newcommand\eeqa{\end{eqnarray}}

\newcommand{\TT}{\widetilde{T}}

\newcommand{\al}{\alpha}

\begin{document}

\title{Kinetic model for transport in granular mixtures}
\author{Pablo Avil\'es}
\affiliation{Departamento de F\'{\i}sica, Universidad de Extremadura, E-06071 Badajoz, Spain}
\author{David Gonz\'alez M\'endez}
\affiliation{Departamento de F\'{\i}sica, Universidad de Extremadura, E-06071 Badajoz, Spain}
\author{Vicente Garz\'{o}}
\email{vicenteg@unex.es; \url{https://fisteor.cms.unex.es/investigadores/vicente-garzo-puertos/}}
\affiliation{Departamento de F\'{\i}sica and Instituto de Computaci\'on Cient\'{\i}fica Avanzada (ICCAEx), Universidad de Extremadura, E-06006 Badajoz, Spain}

\begin{abstract}

A kinetic model for granular mixtures is considered to study three different non-equilibrium situations. The model is based on the equivalence between a gas of elastic hard spheres subjected to a drag force proportional to the particle velocity and a gas of inelastic hard spheres. As a first problem, the relaxation of the velocity moments to their forms in the homogeneous cooling state (HCS) is studied. Then, taking the HCS as the reference state, the kinetic model is solved by the Chapman-Enskog method, which is conveniently adapted to inelastic collisions. For small spatial gradients, the mass, momentum and heat fluxes of the mixture are determined and exact expressions for the Navier-Stokes transport coefficients are obtained. As a third nonequilibrium problem, the kinetic model is solved exactly in the uniform shear flow (USF) state, where the rheological properties of the mixture are computed in terms of the parameter space of the mixture. In addition to the transport properties, the velocity distribution functions of each species are also explicitly obtained. To assess the reliability of the model, its theoretical predictions are compared with both (approximate) analytical results and computer simulations of the original Boltzmann equation. In general, the comparison shows a reasonable agreement between the two kinetic equations. While the diffusion transport coefficients show excellent agreement with the Boltzmann results, more quantitative differences appear in the case of the shear viscosity coefficient and the heat flux transport coefficients. In the case of the USF, although the model qualitatively captures the shear rate dependence of the rheological properties well, the discrepancies increase with increasing inelasticity in collisions.
\end{abstract}

\date{\today}
\maketitle

\section{Introduction}
\label{sec1}

It is now well established that granular media behave like a fluid when externally excited. Under these conditions (rapid flow conditions), granular media can be modeled as a gas of hard spheres with inelastic collisions. In the simplest version, the spheres are assumed to be completely \textit{smooth} and so the inelasticity is accounted for by a (constant) positive coefficient of normal restitution. In the low-density regime, the Boltzmann equation (conveniently generalized to dissipative dynamics) has been used as a starting point to derive the corresponding Navier-Stokes hydrodynamic equations with explicit forms for the transport coefficients. \cite{BP04,G19} On the other hand, as with elastic collisions, \cite{CC70} the determination of the Navier-Stokes transport coefficients requires the solution of a set of coupled linear integral equations. These equations are usually solved by considering the leading terms in a Sonine polynomial expansion. This procedure becomes more tedious in the case of granular mixtures, since not only are the number of transport coefficients greater than for a single component gas, but they also depend on more parameters. \cite{JM87,JM89,Z95,GD02,GMD06,SGNT06,GM07,GDH07,GHD07}

Beyond the Navier-Stokes domain (small spatial hydrodynamic gradients), the computation of transport properties from the Boltzmann equation (for both elastic and/or inelastic collisions) is a very difficult task. For this reason, it is therefore quite common in kinetic theory to resort to alternative approaches for such far from equilibrium states. One possibility is to keep the structure of the (inelastic) Boltzmann collision operator but to assume a different interaction model: the so-called inelastic Maxwell model (IMM). As for the conventional Maxwell molecules, \cite{CC70} IMM's are characterized by the property that the collision rate is independent of the relative velocity of the two colliding spheres. \cite{BK00,BCG00,EB02b} This simplification allows to \textit{exactly} evaluate the moments of the Boltzmann collision operators without an explicit knowledge of the distribution functions. \cite{G03bis,GS07,SG23} However, although the use of IMM's opens up the possibility of obtaining exact results from the Boltzmann equation, these IMM's do not describe real particles, since they do not interact according to a given potential law.

Another possible alternative for obtaining accurate results is to consider a kinetic model equation of the inelastic Boltzmann equation for hard spheres. Kinetic models have proven to be very useful for the analysis of transport properties in far-from-equilibrium states of dilute molecular gases. In fact, for several non-equilibrium situations, exact solutions of the kinetic models have been shown to agree very well with Monte Carlo simulations of the Boltzmann equation for molecular gases. \cite{D90,GS03}  In the case of single-component gases of inelastic hard spheres (IHS), several models have been proposed in the granular literature. \cite{BMD96,BDS99,DBZ04} On the other hand, the number of kinetic models for multicomponent granular mixtures is much smaller. In fact, we are aware of only one kinetic model proposed years ago by Vega Reyes \textit{et al.} \cite{VGS07} This is in contrast to the large number of kinetic models proposed in the literature for molecular mixtures. \cite{GK56,S62,H65,H66,GS67,GSB89,AAP02,HHKPW21,LZW24}  The model reported by Vega Reyes \textit{et al.} \cite{VGS07} is essentially based on the equivalence between a system of elastic hard spheres, subject to a drag force proportional to the particle velocity, and a gas of IHS. \cite{SA05} The relaxation term appearing in the kinetic model can be chosen among the different kinetic models \cite{GK56,S62,H65,H66,GS67,GSB89,AAP02,HHKPW21,LZW24} published in the literature for molecular mixtures of hard spheres. Here, for the sake of simplicity, we have adopted the Gross and Krook (GK) model \cite{GK56} proposed many years ago for studying transport properties in multicomponent molecular gases. Thus, the kinetic model employed in this paper can be considered as a direct extension of the GK model to granular mixtures.

Although the kinetic model of Vega Reyes \textit{et al.} \cite{VGS07} was reported several years ago, to the best of our knowledge it has not been considered so far to study linear and nonlinear transport properties of granular mixtures. The aim of this paper is to consider the above kinetic model to determine the dynamical properties of granular binary mixtures in different non-equilibrium situations. In addition, apart from obtaining the above properties, the simplicity of the model allows one to get the explicit forms of the velocity distribution functions. This is likely one of the main advantages of using a kinetic model instead of the original Boltzmann equation.

Three different but related problems are studied. First, the so-called HCS is analyzed; we are mainly interested here in studying the relaxation of the velocity moments towards their HCS expressions (starting from arbitrary initial conditions). Then, once the HCS is well characterized for the mixture, we solve the kinetic model using the Chapman-Enskog method \cite{CC70} for states close to the HCS. In contrast to the results obtained from the Boltzmann equation, \cite{GD02,SGNT06,GMD06,SNG07,GM07} exact expressions for the complete set of Navier-Stokes transport coefficients of the mixture are derived in terms of the parameter space of the system. Finally, as a third problem, the rheological properties of a binary granular mixture under USF are obtained explicitly.

The search for exact solutions of kinetic models is interesting not only from a formal point of view, but also as a way to assess the reliability of these solutions. To gauge their accuracy, we compare in this paper the theoretical predictions of the kinetic model with (i) (approximate) analytical results of the original Boltzmann equation and with (ii) computer simulation results available in the granular literature. This type of comparison allows us to measure the degree of reliability of the kinetic model for describing granular flows under realistic conditions.

The structure of the paper is as follows. In Sec.\ \ref{sec2} we introduce the original Boltzmann equation for granular mixtures and its balance hydrodynamic equations, and present the explicit form of the kinetic model. Section \ref{sec3} deals with the HCS: a homogeneous state with a granular temperature decaying with time. As said before, we first study the relaxation of the velocity moments to their (steady) asymptotic expressions. It is shown that for certain values of the parameters of the system, quite high velocity moments can diverge in time in the HCS. Section \ref{sec4} is devoted to the application of the Chapman-Enskog method to the kinetic model for obtaining the Navier-Stokes transport coefficients. Their expressions are also compared with both theoretical approximations and computer simulations obtained from the Boltzmann equation. The USF is studied in Sec.\ \ref{sec5} while a brief discussion of the results reported in the paper is given in Sec.\ \ref{sec6}.

\section{Boltzmann kinetic equation for granular mixtures. A kinetic model}
\label{sec2}

We consider an isolated binary granular mixture of inelastic hard spheres of masses $m_i$ and diameters $\sigma_i$ ($i=1,2$). The subscript $i$ labels one of the $s$ mechanically different species or components of the mixture. We assume also for simplicity that the spheres are completely smooth and hence, in a binary collision of particles of the species $i$ with particles of the species $j$ while the
magnitude of the tangential component of the relative velocity of the two colliding spheres remains unaltered, its normal component is reversed and shrunk by a factor $\al_{ij}$. The parameter $\al_{ij}$ ($0<\al_{ij}\leqslant 1$) is called the (constant) coefficient of normal restitution and accounts for the energy dissipated in each binary collision between particles of species $i$ and $j$. In the low-density limit, a kinetic theory description is appropriate, and the one-particle velocity distribution function $f_i(\mathbf{r}, \mathbf{v}, t)$ of species $i$ verifies the set of two-coupled nonlinear integro-differential Boltzmann kinetic equations \cite{G19}
\beq
\label{2.1}
\frac{\partial}{\partial t}f_i+\mathbf{v}\cdot \nabla f_i=\sum_{j=1}^2 J_{ij}[\mathbf{v}|f_i,f_j], \quad i=1,2.
\eeq
where $J_{ij}[f_i,f_j]$ is the inelastic version of the Boltzmann collision operator. Its explicit form can be found for instance in Ref.\ \onlinecite{G19}. At a hydrodynamic level, the relevant fields are the number densities $n_i$, the flow velocity $\mathbf{U}$, and the granular temperature $T$. In terms of moments of the velocity distribution functions $f_i$, they are defined as
\beq
\label{2.2}
n_i=\int d\mathbf{v} f_i(\mathbf{v}),
\eeq
\beq
\label{2.3}
\rho \mathbf{U}=\sum_{i}^2\; m_i n_i \mathbf{U}_i=\sum_{i=1}^2\;m_i \int d\mathbf{v} \mathbf{v}f_i(\mathbf{v}),
\eeq
\beq
\label{2.4}
n T=p=\sum_{i}^2\; n_i T_i =\sum_{i=1}^2\;\frac{m_i}{3} \int d\mathbf{v} \mathbf{V}^2f_i(\mathbf{v}).
\eeq
In Eqs.\ \eqref{2.2}--\eqref{2.4}, $\mathbf{V}=\mathbf{v}-\mathbf{U}$ is the peculiar velocity, $n=\sum_i n_i$ is the total number density, $\rho=\sum_i \rho_i=\sum_i m_i n_i$ is the total mass density, and $p$ is the hydrostatic pressure. Furthermore, the second equality in Eq.\ \eqref{2.3} and the third equality in Eq.\ \eqref{2.4} define the flow velocity $\mathbf{U}_i$ and the kinetic temperature $T_i$ for species $i$, respectively. The partial temperature $T_i$ is a measure of the mean kinetic of energy of particles of species $i$.

The Boltzmann collision operators $J_{ij}[f_i,f_j]$ conserve the number density of each species and the total momentum in each collision $ij$:
\beq
\label{2.5}
\int d\mathbf{v} J_{ij}[\mathbf{v}|f_i,f_j]=0,
\eeq
\beq
\label{2.6}
m_i\int d\mathbf{v} \mathbf{v} J_{ij}[\mathbf{v}|f_i,f_j]+m_j\int d\mathbf{v} \mathbf{v} J_{ji}[\mathbf{v}|f_j,f_i]=0.
\eeq
Nevertheless, unless $\al_{ij}=1$, the operators $J_{ij}[f_i,f_j]$ do not conserve the kinetic energy in each collision $ij$:
\beq
\label{2.7}
m_i\int d\mathbf{v} \mathbf{v}^2 J_{ij}[\mathbf{v}|f_i,f_j]+m_j\int d\mathbf{v} \mathbf{v}^2 J_{ji}[\mathbf{v}|f_j,f_i]\neq 0.
\eeq
The total cooling rate $\zeta$ due to collisions among all species is given by
\beq
\label{2.8}
\zeta=-\frac{1}{3nT}\sum_{i=1}^2\sum_{j=1}^2\; \int d\mathbf{v} m_i v^2 J_{ij}[\mathbf{v}|f_i,f_j].
\eeq

The corresponding balance hydrodynamic equations for the densities of mass, momentum, and kinetic energy can easily be derived from the properties \eqref{2.5}--\eqref{2.8} of the Boltzmann collision operators $J_{ij}[f_i,f_j]$. They are given by \cite{G19}
\beq
\label{2.9}
D_t n_i+n_i\nabla\cdot \mathbf{U}+\frac{\nabla\cdot\mathbf{j}_i}{m_i}=0,
\eeq
\beq
\label{2.10}
D_t\mathbf{U}+\rho^{-1}\nabla\cdot\mathsf{P}=\mathbf{0},
\eeq
\beq
\label{2.11}
D_tT-\frac{T}{n}\sum_{i=1}^s\frac{\nabla\cdot\mathbf{j}_i}{m_i}+\frac{2}{3n}
\left(\nabla\cdot\mathbf{q}+\mathsf{P}:\nabla\mathbf{U}\right)=-\zeta T.
\eeq
In Eqs.\ \eqref{2.9}--\eqref{2.11},
\beq
\label{2.12}
\mathbf{j}_i=m_i\int\;d\mathbf{v}\; \mathbf{V}f_i(\mathbf{v})
\eeq
is the mass flux for component $i$ relative to the local flow $\mathbf{U}$,
\beq
\label{2.13}
\mathsf{P}=\sum_{i=1}^2 \mathsf{P}_i=\sum_{i=1}^2\int d\mathbf{v}\; m_i\mathbf{V}\mathbf{V}f_i(\mathbf{v})
\eeq
is the (total) pressure tensor and,
\beq
\label{2.14}
\mathbf{q}=\sum_{i=1}^2 \mathbf{q}_i=\sum_{i=1}^2\int d\mathbf{v}\; \frac{m_i}{2}V^2\mathbf{V}f_i(\mathbf{v})
\eeq
is the (total) heat flux. The first equality in Eqs.\ \eqref{2.13} and \eqref{2.14} defines the partial contributions $\mathsf{P}_i$ and $\mathbf{q}_i$ to the pressure tensor and the heat flux, respectively. A consequence of the definition \eqref{2.12} is that $\mathbf{j}_1=-\mathbf{j}_2$. In addition, for the sake of simplicity, we take the Boltzmann constant $k_\text{B}=1$ throughout the paper.

It is quite obvious that the hydrodynamic equations \eqref{2.9}--\eqref{2.11} are not a closed set of differential equations for the hydrodynamic fields $n_i$, $\mathbf{U}$ and $T$. This can be achieved by expressing the fluxes and the cooling rate as functions of the hydrodynamic fields and their gradients (constitutive equations). To obtain these equations, for small spatial gradients, one can solve the Boltzmann equation \eqref{2.1} using the Chapman-Enskog method \cite{CC70} conveniently adapted to dissipative collisions. For IHS, approximate forms for the Navier-Stokes transport coefficients have been derived by considering the lowest Sonine approximation. \cite{GD02,SGNT06,GMD06,GM07}

Given the difficulties associated with the complex mathematical structure of the Boltzmann collision operators $J_{ij}[f_i,f_j]$ for IHS, a possible way to overcome them while preserving the structure of the above operators is to consider the so-called IMM. \cite{BK00,BCG00,CCG00,BK03} As for elastic Maxwell models, \cite{TM80} the collision rate for IMM does not depend of the relative velocity of the colliding spheres, and so one can evaluate exactly the collision moments of $J_{ij}[f_i,f_j]$ without explicit knowledge of the distributions $f_i$ and $f_j$. This property allows an exact determination of the Navier-Stokes transport coefficients \cite{GA05} as well as the rheological properties of a sheared granular mixture. \cite{G03bis} However, despite their practical usefulness, these IMM's do not interact according to a given potential law and can be considered as a toy model for unveiling the influence of dissipation on transport in granular flows.
As an alternative to IMM for obtaining accurate results in granular mixtures of IHS, one can consider kinetic models.

\subsection{Kinetic model for granular mixtures}

As mentioned in Sec.\ \ref{sec1}, the idea behind the construction of a kinetic model is to replace the operator $J_{ij}[f_i,f_j]$ for IHS by a simpler mathematical
collision term that retains its relevant physical properties. While in the case of molecular mixtures many different kinetic models
have been proposed in the literature, \cite{GK56,S62,H65,H66,GS67,GSB89,AAP02,HHKPW21,LZW24}
kinetic models for granular mixtures are much more scarce. To the best of our knowledge, only one kinetic model has been reported in the granular literature: the model proposed years ago by Vega Reyes \emph{et al.} \cite{VGS07} This model is essentially based on the equivalence between a system of elastic spheres subject to a drag force proportional to the (peculiar) velocity $\mathbf{v}-\mathbf{U}_i$ with a gas of IHS. \cite{SA05} According to this equivalence, the Boltzmann collision operator $J_{ij}[f_i,f_j]$ is replaced by the term \cite{VGS07}
\beq
\label{2.15}
Q_{ij}[\mathbf{v}|f_i,f_j]=\xi_{ij}K_{ij}[\mathbf{v}|f_i,f_j]+\frac{\epsilon_{ij}}{2}\frac{\partial}{\partial {\bf v}}\cdot ({\bf v}-{\bf U}_i)f_i.
\eeq
While the quantities $\xi_{ij}$ and $\epsilon_{ij}$ are determined by optimizing the agreement between the kinetic model and the Boltzmann equation, the term $K_{ij}[\mathbf{v}|f_i,f_j]$ can be modeled as a simple relaxation term, which can be chosen from among the various kinetic models proposed in the literature for molecular (elastic) mixtures. \cite{GK56,S62,H65,H66,GS67,GSB89,AAP02,HHKPW21,LZW24}  It is quite obvious from Eq.\ \eqref{2.15} that the quantity $\epsilon_{ij}\geq 0$ can be regarded as the coefficient of the drag (friction) force $\mathbf{F}_{ij}=-(m_i\epsilon_{ij}/2)(\mathbf{v}-\mathbf{U}_i)$ felt by the (elastic) particles of species $i$. The main goal of this non-conservative force is to mimic the loss of energy that occurs in a granular mixture when particles of species $i$ collide with particles of species $j$.

The parameters $\xi_{ij}$ and $\epsilon_{ij}$ of the model are determined by requiring that the collisional transfer of momentum and energy of species $i$ due to collisions with particles of species $j$ must be the same as those obtained from the Boltzmann kinetic equation. Given that these later collisional moments cannot be exactly obtained, one replaces the true velocity distributions $f_i$ by their Maxwellian forms
\beq
\label{2.15.0}
f_{i,\text{M}}(\mathbf{v})=n_i\left(\frac{m_i}{2\pi \TT_i}\right)^{3/2} \exp{\left[-\frac{m_i}{2\TT_i}(\mathbf{v}-\mathbf{U}_i)^2\right]},
\eeq
where
\beq
\label{2.18}
\TT_i=\frac{m_i}{3n_i}\int d\mathbf{v}\; (\mathbf{v}-\mathbf{U}_i)^2 f_i(\mathbf{v})=T_i-\frac{m_i}{3}\left(\mathbf{U}_i-\mathbf{U}\right)^2.
\eeq
By using the Maxwellian approximation \eqref{2.15.0}, $\xi_{ij}$ is simply given by
\beq
\label{2.15.1}
\xi_{ij}=\frac{1+\al_{ij}}{2},
\eeq
while $\epsilon_{ij}$ is \cite{VGS07}
\beq
\label{2.16}
\epsilon_{ij}=\frac{1}{2}\nu_{ij}\mu_{ji}^2\left[1+\frac{m_i\widetilde{T}_j}{m_j
\widetilde{T}_i}+\frac{3}{6}\frac{m_i}{\widetilde{T}_i}\left({\bf
U}_i-{\bf U}_j\right)^2\right](1-\alpha_{ij}^2).
\eeq
Here,
\beq
\label{2.17}
\nu_{ij}=\frac{8\sqrt{\pi}}{3}n_j\sigma_{ij}^{2}\left(\frac{2\widetilde{T}_i}{m_i}+
\frac{2\widetilde{T}_j}{m_j}\right)^{1/2}
\eeq
is an effective collision frequency for IHS and
$\sigma_{ij}=(\sigma_i+\sigma_j)/2$.

\vicente{It is quite obvious that the use of the Maxwellian approximation to estimate the parameters $\xi_{ij}$ and $\epsilon_{ij}$ could introduce potential deviations of the results derived from the kinetic model from those obtained from the original Boltzmann equation. These differences could be partially mitigated by considering the first few terms in the Sonine polynomial expansion of $f_i$. However, although the inclusion of these non-Gaussian corrections to $\xi_{ij}$ and $\epsilon_{ij}$ could reduce the discrepancies between the kinetic model and the Boltzmann equation for strong dissipation, it would result in a rather complicated kinetic model. Therefore, for practical purposes, it is more desirable to estimate $\xi_{ij}$ and $\epsilon_{ij}$ using the Maxwellian approximation \eqref{2.15.0}. As will be shown later, the accuracy of the estimates \eqref{2.15.1} and \eqref{2.16} is justified, for example, by the excellent agreement found between the theoretical predictions of the temperature ratio $T_1/T_2$ obtained from the kinetic model in the HCS and computer simulations (see figure \ref{fig4}).

To complete the definition of the kinetic model, it remains to choose the form of the term $K_{ij}[f_i,f_j]$.
As mentioned in section \ref{sec1}, $K_{ij}[f_i,f_j]$ is chosen as the relaxation term}
\beq
\label{2.17.1}
K_{ij}[f_i,f_j]=-\nu_{ij}(f_i-f_{ij}),
\eeq
where the form of the reference distribution $f_{ij}$ is provided by the kinetic model for gas mixtures proposed by Gross and Krook: \cite{GK56}
\beq
\label{2.19}
f_{ij}(\mathbf{v})=n_i\left(\frac{m_i}{2\pi
T_{ij}}\right)^{3/2} \exp\left[-\frac{m_i}{2T_{ij}}\left(\mathbf{v}-\mathbf{U}_{ij}\right)^2\right].
\eeq
In Eq.\ \eqref{2.19}, we have introduced the quantities
\beq
\label{2.20}
\mathbf{U}_{ij}=\frac{m_i}{m_i+m_j}\mathbf{U}_i+\frac{m_j}{m_i+m_j}\mathbf{U}_j,
\eeq
\begin{widetext}
\beq
\label{2.21}
T_{ij}=\TT_i+\frac{2m_i
m_j}{(m_i+m_j)^2}\Bigg\{\TT_j-\TT_i+\frac{(\mathbf{U}_i-\mathbf{U}_j)^2}{6}
\left[m_j+
\frac{\TT_j-\TT_i}{\TT_i/m_i+\TT_j/m_j}\right]\Bigg\}.
\eeq

In summary, the kinetic model for a low-density granular binary mixture of IHS (which can be seen as the natural extension of the GK model to granular mixtures) is given by
\begin{equation}
\label{2.22}
\partial_tf_i+{\bf v}\cdot \nabla f_i=-\sum_{j=1}^2 \frac{1+\alpha_{ij}}{2}\nu_{ij}\left(f_i-f_{ij}\right)
+\sum_{j=1}^2\frac{\epsilon_{ij}}{2}
\frac{\partial}{\partial {\bf v}}\cdot ({\bf v}-{\bf U}_i)f_i,\quad i=1,2,
\end{equation}
where $\epsilon_{ij}$, $\nu_{ij}$, and $f_{ij}$ are defined by Eqs.\ \eqref{2.16}, \eqref{2.17}, and \eqref{2.19}, respectively.
The kinetic model \eqref{2.2} is the starting point to analyze different nonequilibrium problems. This study will be carried out in the next three sections.
\end{widetext}

\section{Homogeneous cooling state}
\label{sec3}

We assume that the granular binary mixture is in a spatially homogeneous state. In contrast to the (conventional) molecular mixtures of hard spheres, the mixture does not evolve towards an equilibrium state
characterized by the Maxwellian distribution \eqref{2.15.0} with $\mathbf{U}_i=\mathbf{0}$ and $T_i=T$.
This is because the Maxwellian distributions are not solutions of the inelastic version of the homogeneous set of Boltzmann equations. On the other hand, if one assumes homogenous initial conditions, after a few collision times the mixture reaches a special hydrodynamic state: the so-called HCS. \cite{NE98,GD99b} In the HCS, the granular temperature $T(t)$ monotonically decays in time. In this case, without loss generality, $\mathbf{U}_1=\mathbf{U}_2=\mathbf{0}$ and hence the set of kinetic equations \eqref{2.22} for $f_1$ and $f_2$ becomes
\beq
\label{3.1}
\partial_t f_1=-\frac{1}{2}\omega_1 f_1+\frac{\omega_{11}f_{11}+\omega_{12}f_{12}}{2}+\frac{\epsilon_{1}}{2}\frac{\partial}{\partial \mathbf{v}}\cdot \mathbf{v}f_1,
\eeq
\beq
\label{3.2}
\partial_t f_2=-\frac{1}{2}\omega_2 f_2+\frac{\omega_{22}f_{22}+\omega_{21}f_{21}}{2}+\frac{\epsilon_{2}}{2}\frac{\partial}{\partial \mathbf{v}}\cdot \mathbf{v}f_2.
\eeq
In Eqs.\ \eqref{3.1} and \eqref{3.2},  $\omega_1=\omega_{11}+\omega_{12}$, $\omega_2=\omega_{22}+\omega_{21}$, $\epsilon_1=\epsilon_{11}+\epsilon_{12}$, $\epsilon_2=\epsilon_{22}+\epsilon_{21}$, and
\beq
\label{3.3}
\omega_{ij}=(1+\alpha_{ij})\nu_{ij}, \quad \nu_{ij}=\frac{8\sqrt{\pi}}{3}n_j\sigma_{ij}^{2}\left(\frac{2 T_i}{m_i}+
\frac{2 T_j}{m_j}\right)^{1/2},
\eeq
\beq
\label{3.4}
\epsilon_{ij}=\frac{1}{2}\nu_{ij}\mu_{ji}^2\left(1+\frac{m_i{T}_j}{m_j
{T}_i}\right)(1-\alpha_{ij}^2).
\eeq
In addition, in the HCS the quantities $T_{ij}$ are given by
\beq
\label{3.4.1}
T_{ij}=T_i+\frac{2m_i
m_j}{(m_i+m_j)^2}\left(T_j-T_i\right).
\eeq

For homogeneous states, the mass and heat fluxes vanish ($\mathbf{j}_i=\mathbf{q}=\mathbf{0}$) while the pressure tensor $P_{k\ell}=p\delta_{k\ell}$ where $p= n T$ is the hydrostatic pressure. Thus, the balance equations \eqref{2.9} and \eqref{2.10} trivially hold and the balance equation \eqref{2.11} of the granular temperature  yields
\beq
\label{3.5.1}
\frac{\partial T}{\partial t}=-T \zeta,
\eeq
where the cooling rate $\zeta$ is
\beq
\label{3.6.1}
\zeta=\sum_{i=1}^2 x_i \gamma_i \zeta_i.
\eeq
\vicente{The definition of the partial cooling rates $\zeta_i$ can be easily obtained from Eqs.\ \eqref{2.8} and \eqref{3.6.1} as
\beq
\label{3.6.2bis}
\zeta_i=-\frac{1}{3n_iT_i}\sum_{j=1}^2\int d\mathbf{v}\; m_i v^2 J_{ij}[f_i,f_j].
\eeq
Within the context of the kinetic model \eqref{2.22}, the operator $J_{ij}[f_i,f_j]$ is replaced by the term
\beq
\label{3.6.3}
J_{ij}[f_i,f_j]\to -\frac{1+\al_{ij}}{2}\nu_{ij}\left(f_i-f_{ij}\right)+\frac{\epsilon_{ij}}{2}\frac{\partial}{\partial {\bf v}}\cdot \mathbf {v}f_i.
\eeq
Substitution of Eq.\ \eqref{3.6.3} into Eq.\ \eqref{3.6.2bis} allows us to exactly evaluate the partial cooling rates $\zeta_i$. They are given by}
\beqa
\label{3.6.2}
\zeta_i&=&\sum_{j=1}^2 \nu_{ij}\frac{m_im_j}{(m_i+m_j)^2}
(1+\al_{ij})\Bigg[\frac{T_i-T_j}{T_i}
\nonumber\\
& & +\frac{1-\al_{ij}}{2} \Bigg(
\frac{m_j}{m_i}+\frac{T_j}{T_i}\Bigg)\Bigg],
\eeqa
where $x_i=n_i/n$ is the concentration or mole fraction of species $i$ and $\gamma_i=T_i/T$ is the temperature ratio of species $i$. It must be remarked that the expression \eqref{3.6.1} for the cooling rate coincides with the one obtained from the original Boltzmann equation \cite{GD99b} when one approaches the distributions $f_i$ by their Maxwellian forms \eqref{2.15.0}.

At a kinetic level, it is also interesting to analyze the time evolution of the partial temperatures $T_i$. From Eqs.\ \eqref{3.1} and \eqref{3.2}, one easily gets
\beq
\label{3.5}
\frac{\partial T_1}{\partial t}=-T_1 \zeta_1, \quad \frac{\partial T_2}{\partial t}=-T_2 \zeta_2,
\eeq
where the cooling rates $\zeta_i$ are given by Eq.\ \eqref{3.6.2}. The time evolution of the temperature ratio $\gamma(t)=T_1(t)/T_2(t)$ follows from Eq.\ \eqref{3.5} as
\beq
\label{3.6}
\frac{\partial \gamma}{\partial t}=\gamma(\zeta_2-\zeta_1).
\eeq
As computer simulations clearly show, \cite{MG02,DHGD02} after a transient period, the granular mixture reaches a hydrodynamic regime where the time dependence of the distributions $f_i$ is only through their dependence on the (global) granular temperature $T(t)$. This implies that the temperature ratio $\gamma$ is \emph{independent} of time. However, in contrast to molecular (elastic) mixtures, $\gamma\neq 1$ and so, in general the total kinetic energy of the mixture is not equally distributed between both species (breakdown of energy equipartition). Results derived from kinetic theory, \cite{GD99b} computer simulations, \cite{MG02,BT02,DHGD02,PMP02,BT02b,CH02,KT03,WJM03,BRM05,SUKSS06} and even real experiments in driven \cite{WP02,FM02} and freely cooling mixtures \cite{PTHS24} have clearly shown that the temperature ratio $T_1(t)/T_2(t)$ is in general different from 1; it exhibits in fact a complex dependence on the parameter space of the mixture. Since the temperature ratio $\gamma$ reaches a steady value in the HCS, then according to Eq.\ \eqref{3.6} the partial cooling rates must be equal:
\beq
\label{3.7}
\zeta_1=\zeta_2.
\eeq
The numerical solution to the condition \eqref{3.7} provides the dependence of $\gamma$ on the parameters of the binary granular mixture.

Regarding the distribution functions $f_i(\mathbf{v};t)$, dimensional analysis shows that in the HCS these distributions adopt the form
\beq
\label{3.8}
f_i(\mathbf{v};t)=n_i v_\text{th}^{-3}(t)\varphi_i\left(\mathbf{c}\right),
\eeq
where $\mathbf{c}=\mathbf{v}/v_\text{th}(t)$, $v_\text{th}(t)=\sqrt{2T(t)/\overline{m}}$ is a thermal velocity defined in terms of the (global) granular temperature $T(t)$ and $\overline{m}=(m_1+m_2)/2$. In the context of the original Boltzmann equation, the explicit form of the scaled distribution $\varphi_i$ is not yet known. Approximate expressions for this distribution \cite{G19} can be obtained by truncating the Sonine polynomial expansion of $\varphi_i$. On the other hand, the use of the kinetic model allows us to provide an exact form of the scaled distributions $\varphi_i$ in the HCS. This is done in the subsection \ref{sec3.C}. The possibility of obtaining the exact form of $\varphi_i$ is probably one of the major advantages of considering a kinetic model instead of the true Boltzmann equation.

\subsection{Relaxation of the velocity moments toward their HCS forms}
\label{sec3.A}

Apart from the partial temperatures, it is worthwhile studying the time evolution of the high-degree velocity moments. To do it, let us introduce the canonical moments
\beq
\label{3.9}
M_{k_1,k_2,k_3}^{(i)}(t)=\int d\mathbf{v}\; v_x^{k_1}v_y^{k_2}v_z^{k_3}f_i(\mathbf{v};t), \quad (i=1,2).
\eeq
The time evolution of these moments can be easily derived when one multiplies both sides of Eqs.\ \eqref{3.1} and \eqref{3.2} by $v_x^{k_1}v_y^{k_2}v_z^{k_3}$ and integrates over velocity. The result is
\beqa
\label{3.10}
\partial_t M_{k_1,k_2,k_3}^{(1)}&+&\frac{\omega_1+k \epsilon_1}{2}M_{k_1,k_2,k_3}^{(1)}=\frac{1}{2}n_1 v_{\text{th}}^k \Big(\omega_{11}\theta_1^{-k/2}
\nonumber\\
& &
+\omega_{12}\theta_{12}^{-k/2}\Big)\Gamma_{k_1k_2k_3},
\eeqa
where $k=k_1+k_2+k_3$ is the degree of the moment, $\theta_1=m_1T/(\overline{m}T_1)$, and $\theta_{12}=m_1T/(\overline{m}T_{12})$. Moreover, in Eq.\ \eqref{3.10}, we have introduced the shorthand notation
\beq
\label{3.11}
\Gamma_{k_1k_2k_3}\equiv \pi^{-3/2}\Gamma\left(\frac{k_1+1}{2}\right)\Gamma\left(\frac{k_2+1}{2}\right)\Gamma\left(\frac{k_3+1}{2}\right)
\eeq
if $k_1$, $k_2$, and $k_3$ are even, being zero otherwise. The time evolution equation of the moments $M_{k_1,k_2,k_3}^{(2)}(t)$ for the species 2 can be easily inferred from Eq.\ \eqref{3.10} by making the change $1\leftrightarrow 2$.

It is convenient to introduce the dimensionless velocity moments
\beq
\label{3.12}
M_{k_1,k_2,k_3}^{*(i)}(t)=n_i^{-1}v_{\text{th}}^{-k}M_{k_1,k_2,k_3}^{(i)}(t).
\eeq
In the HCS one expects that after a transient regime the dimensionless moments $M_{k_1,k_2,k_3}^{*(i)}(t)$ reach an asymptotic steady value. The time evolution of the dimensionless moments $M_{k_1,k_2,k_3}^{*(1)}(t)$ is obtained from Eq.\ \eqref{3.10} when one takes into account the time evolution equation \eqref{3.6} for the granular temperature $T(t)$. It can be written as
\beqa
\label{3.13}
\partial_\tau M_{k_1,k_2,k_3}^{*(1)}&+&\frac{\omega_1^*+k(\epsilon_1^*-\zeta^*)}{2}M_{k_1,k_2,k_3}^{*(1)}=
\frac{1}{2}\Big(\omega_{11}^*\nonumber\\
& &\times \theta_1^{-k/2} +\omega_{12}^*\theta_{12}^{-k/2}\Big)\Gamma_{k_1k_2k_3},
\eeqa
where $\omega_{ij}^*=\omega_{ij}/\nu$, $\omega_{1}^*=\omega_{1}/\nu$, $\epsilon_{1}^*=\epsilon_{1}/\nu$, and $\zeta^*=\zeta/\nu$. Furthermore, $\nu(t)=n \sigma_{12}^2 v_\text{th}(t)$ is an effective collision frequency and $\tau$ is the dimensionless time
\beq
\label{3.14}
\tau=\int_0^t\; ds \; \nu(s).
\eeq
The parameter $\tau$ measures time as the number of (effective) collisions per particle.  The solution to Eq.\ \eqref{3.13} is
\beqa
\label{3.15}
M_{k_1,k_2,k_3}^{*(1)}(\tau)&=&\left[M_{k_1,k_2,k_3}^{*(1)}(0)-M_{k_1,k_2,k_3}^{*(1)}(\infty)\right]e^{-\lambda_1^*\tau}
\nonumber\\
& &
+
M_{k_1,k_2,k_3}^{*(1)}(\infty).
\eeqa
Here, the eigenvalue $\lambda_1^*$ is
\beqa
\label{3.17}
\lambda_1^*&=&\frac{\omega_1^*+k(\epsilon_1^*-\zeta^*)}{2}\nonumber\\
&=&\frac{8\sqrt{\pi}}{3}x_1 \left(\frac{\sigma_1}{\sigma_{12}}\right)^2
\sqrt{\frac{2}{\theta_1}}\frac{1+\al_{11}}{2}+\frac{8\sqrt{\pi}}{3}
x_2 \frac{1+\al_{12}}{2}\nonumber\\
& \times& \left(\frac{\theta_1+\theta_2}{\theta_1\theta_2}\right)^{1/2}\Bigg(1-k
\frac{m_1m_2}{(m_1+m_2)^2}\frac{T_1-T_2}{T_1}\Bigg),\nonumber\\
\eeqa
while the asymptotic steady value $M_{k_1,k_2,k_3}^{*(1)}(\infty)$ is
\beq
\label{3.16}
M_{k_1,k_2,k_3}^{*(1)}(\infty)=\frac{\omega_{11}^*\theta_1^{-k/2}+\omega_{12}^*\theta_{12}^{-k/2}
}{2\lambda_1^*}\Gamma_{k_1k_2k_3}.
\eeq
As mentioned before, the corresponding equation for $M_{k_1,k_2,k_3}^{*(2)}(\tau)$ can easily be obtained from the change of $1\leftrightarrow 2$. Since $\gamma_2=(1-x_1\gamma_1)/x_2$, the evolution equation of the moment $M_{k_1,k_2,k_3}^{*(1)}$ is decoupled from that of $M_{k_1,k_2,k_3}^{*(2)}$. This is in contrast to results derived from the original Boltzmann equation, where the moments of species 1 and 2 are coupled in their corresponding time evolution equations. \cite{SG23}

According to Eq.\ \eqref{3.15}, the (scaled) moments of degree $k$ of species $M_{k_1,k_2,k_3}^{*(1)}$ tend asymptotically towards their finite values $M_{k_1,k_2,k_3}^{*(1)}(\infty)$ if the corresponding eigenvalues $\lambda_1^*>0$. For elastic collisions ($\al_{ij}=1$), $T_1=T_2=T$, and Eq.\eqref{3.17} leads to the following expression of $\lambda_1^*$:
\beq
\label{3.16.1}
\lambda_{1,{\text{el}}}^*=\frac{8}{3}\sqrt{\frac{\pi}{\mu_{12}}}\Bigg[x_1\left(\frac{\sigma_1}{\sigma_{12}}\right)^2+\frac{x_2}{\sqrt{2\mu_{21}}} \Bigg]>0.
\eeq
Thus, for molecular mixtures of hard spheres, all velocity moments converge towards their equilibrium values as expected. On the other hand, for granular mixtures ($\al_{ij}\neq 1$), a systematic analysis of the dependence of the eigenvalues $\lambda_1^*$ on the parameter space of the mixture shows that for sufficiently high degree moments, $\lambda_1^*$ can be negative for values of $\al$ smaller than a certain critical value $\al_c$. This means that the moments $M_{k_1,k_2,k_3}^{*(1)}$ diverge in time for $\al<\al_c$. The possibility that higher velocity moments in the HCS may diverge in time in certain regions of the mixture parameter space has also been found in the case of IMM. \cite{SG23}

\begin{figure}[h]
\includegraphics[width=0.4\textwidth]{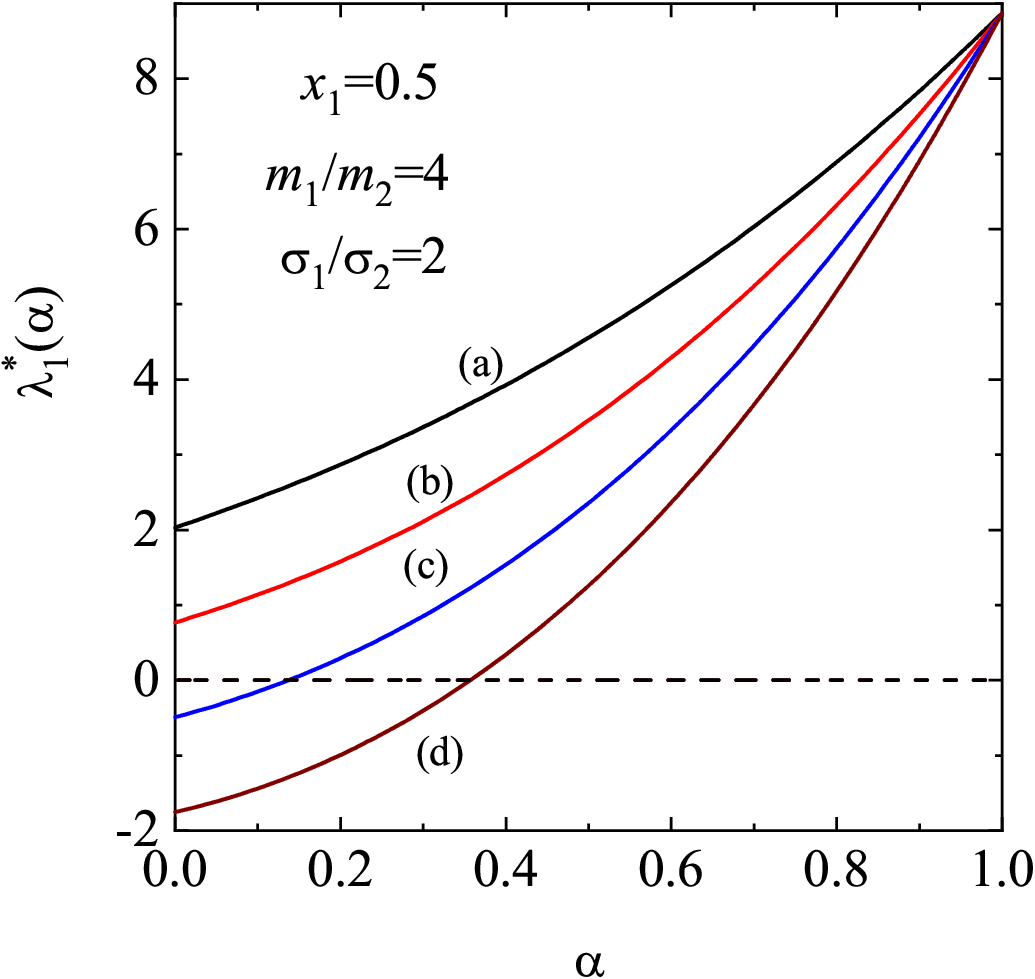}
\caption{Plot of the eigenvalue $\lambda_1^*$ versus the (common) coefficient of restitution $\al$ for an equimolar mixture ($x_1=\frac{1}{2}$) with $\sigma_1/\sigma_2=2$, $m_1/m_2=4$, and four different values of the degree $k$: $k=20$ (a), $k=30$ (b), $k=40$ (c), and $k=50$ (d). The eigenvalue $\lambda_1^*$ is defined by Eq.\  \eqref{3.17}.}
\label{fig1}
\end{figure}

It is quite apparent that a full study of the dependence of the eigenvalue $\lambda_1^*$ on the parameters of the mixture is quite difficult due to the many parameters involved in the problem: $\left(\al_{11}, \al_{22}, \al_{12}, m_1/m_2, \sigma_1/\sigma_2, x_1\right)$. Thus, for the sake of concreteness, we will consider equimolar mixtures ($x_1=1/2$) with a common coefficient of restitution ($\al_{ij}\equiv \al$). To illustrate the $\al$-dependence of $\lambda_1^*$, Fig.\ \ref{fig1} shows $\lambda_1^*(\al)$ for an equimolar mixture ($x_1=\frac{1}{2}$) with $\sigma_1/\sigma_2=2$, $m_1/m_2=4$, and different values of the degree $k=k_1+k_2+k_3$ of the velocity moments $M_{k_1,k_2,k_3}^{*(1)}$. Figure \ref{fig1} highlights that for $k=40$ and 50, $\lambda_1^*(\al)$ becomes negative for $\al<\al_c$. For the mixture considered in Fig.\ \ref{fig1}, $\al_c\simeq 0.144$ for $k=40$ and $\al_c\simeq 0.351$ for $k=50$. This means that, if $\al<\al_c$, the moments $M_{k_1,k_2,k_3}^{*(1)}$ of degree 40 and 50 grow exponentially in time.

\begin{figure}[h]
\includegraphics[width=0.4\textwidth]{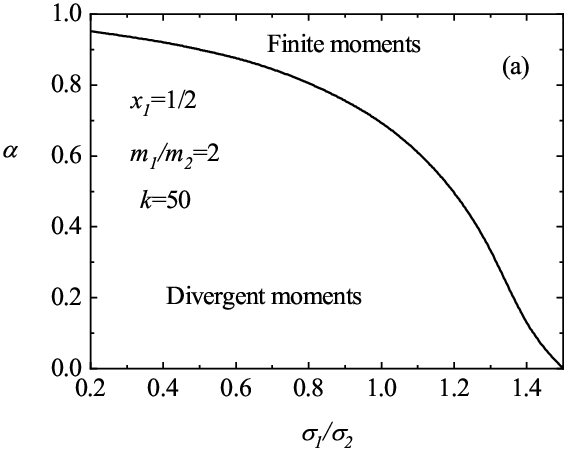}
\includegraphics[width=0.4 \textwidth,angle=0]{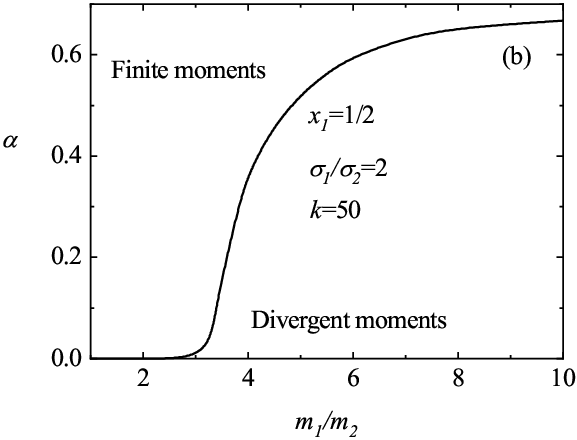}
\caption{Panel (a): Phase diagram in the $(\al, \sigma_1/\sigma_2)$-plane for the asymptotic long time behavior of the moments $M_{k_1,k_2,k_3}^{*(1)}$ with degree $k=50$. Here, $x_1=\frac{1}{2}$ and $m_1/m_2=2$. Panel (b): Phase diagram in the $(\al, m_1/m_2)$-plane for the asymptotic long time behavior of the moments $M_{k_1,k_2,k_3}^{*(1)}$ with degree $k=50$. Here, $x_1=\frac{1}{2}$ and $\sigma_1/\sigma_2=2$.}
\label{fig2}
\end{figure}

To complement Fig.\ \ref{fig1}, panels (a) and (b) of Fig.\ \ref{fig2} show phase diagrams associated with the singular behavior of the moments $M_{k_1,k_2,k_3}^{*(1)}$ of degree 50. In panel (a), $x_1=\frac{1}{2}$ and $m_1/m_2=2$, while in panel (b), $x_1=\frac{1}{2}$ and $\sigma_1/\sigma_2=2$. The curve $\al_c(\sigma_1/\sigma_2)$ ($\al_c(m_1/m_2)$) divides the parameter space of panel (a) (panel (b)) into two regions: The region above the curve corresponds to values of $(\al,\sigma_1/\sigma_2)$ ($(\al,m_1/m_2))$ where these moments are convergent (and thus go to the stationary value $M_{k_1,k_2,k_3}^{*(1)}(\infty)$). Otherwise, the region below the above curves defines states where these moments are divergent. Panel (a) of Fig.\ \ref{fig2} shows that the region of divergent moments grows as the size of the heavier species decreases, while panel (b) highlights the growth of the divergent region as the larger species becomes heavier.

As mentioned above, a similar behavior of the high-velocity moments of the IMM in the HCS has recently been found. \cite{SG23} However, in the special case of IMM, the third-degree velocity moments could already diverge in certain regions of the parameter space of the system. This contrasts with the results derived here, since one has to consider very high degree moments to find such divergences. One might think that this singular behavior could be associated with an algebraic velocity tail in the long time of the distribution function $f_1(\mathbf{v})$ (as in the case of the true Boltzmann equation. \cite{MG02}) However, as we will show later in subsection \ref{sec3.C}, this is not the case, since the form of the distribution function obtained from an exact solution of the kinetic model behaves well for any value of the velocity particle. It could also be possible that this singular behavior is an artifact of the kinetic model, since it generally appears for very high degree moments. Beyond this drawback of the model, one could argue that this unphysical behavior could be related to the absence of the HCS solution \eqref{3.8} for values of the coefficient of restitution smaller than $\al_c$. Clarification of this point requires further analysis; computer simulations of the original Boltzmann equation for high-degree velocity moments may shed light on this issue.

\subsection{Temperature ratio and fourth-degree moments in the HCS}
\label{sec3.B}

\begin{figure}[h]
\includegraphics[width=0.4\textwidth]{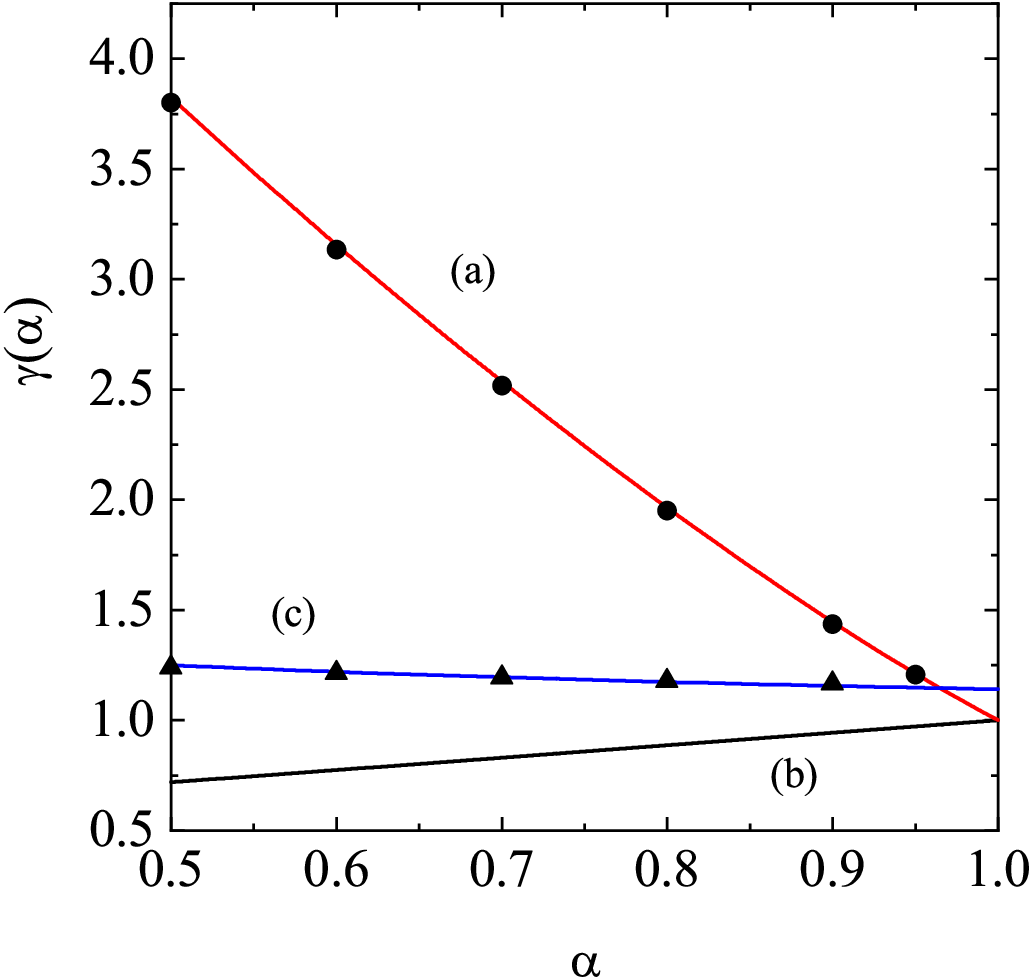}
\caption{Plot of the temperature ratio $\gamma=T_1/T_2$ versus the coefficient of restitution $\al$ for three different mixtures:
$x_1=\frac{2}{3}$, $m_1/m_2=10$, $\sigma_1/\sigma_2=1$, and a common coefficient of restitution $\al_{ij}=\al$ (a); $x_1=\frac{2}{3}$, $m_1/m_2=0.5$, $\sigma_1/\sigma_2=1$, and a common coefficient of restitution $\al_{ij}=\al$ (b); $x_1=\frac{1}{2}$, $m_1/m_2=\sigma_1/\sigma_2=1$, $\al_{11}=0.9$, $\al_{22}=0.5$ and $\al_{12}=\al$ (c). The solid lines are the results obtained from the kinetic model while the symbols refer to Monte Carlo simulations (circles for the case (a) and triangles for the case (c)).}
\label{fig4}
\end{figure}
\begin{figure}[h]
\includegraphics[width=0.41\textwidth]{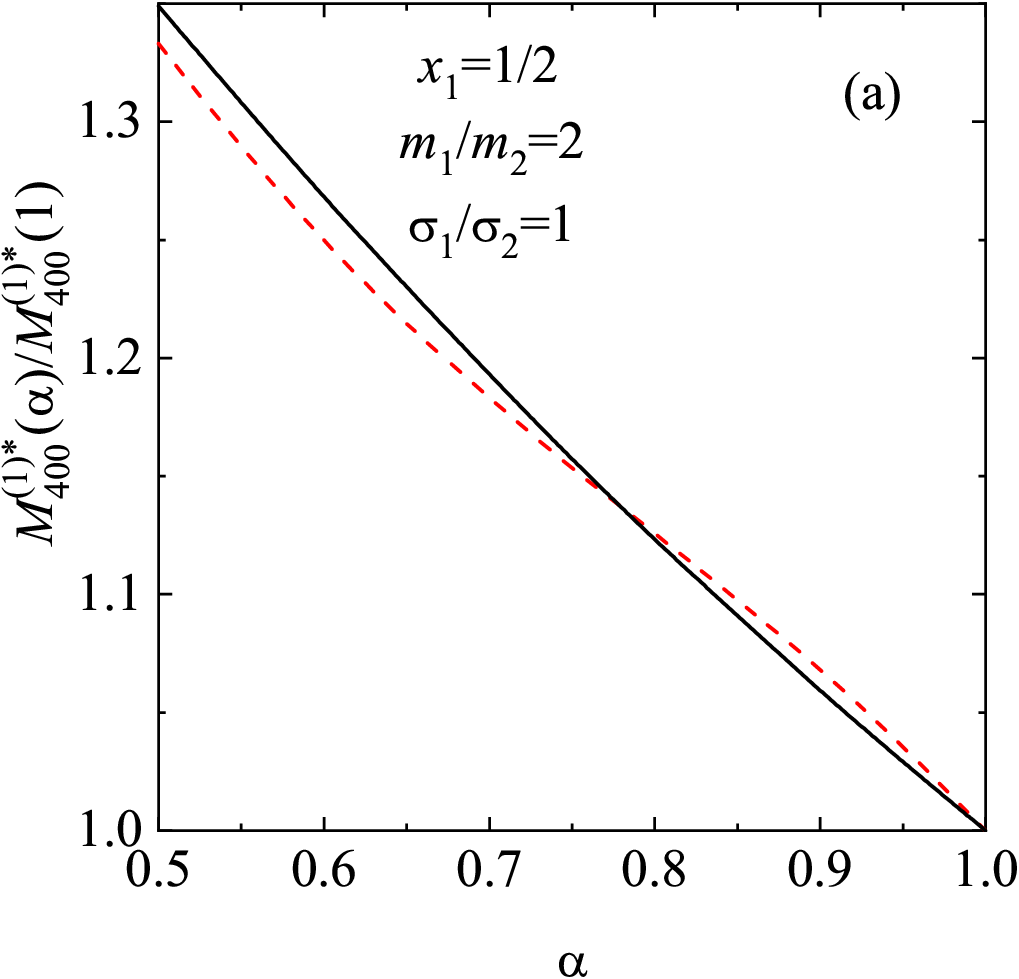}
\includegraphics[width=0.42 \textwidth,angle=0]{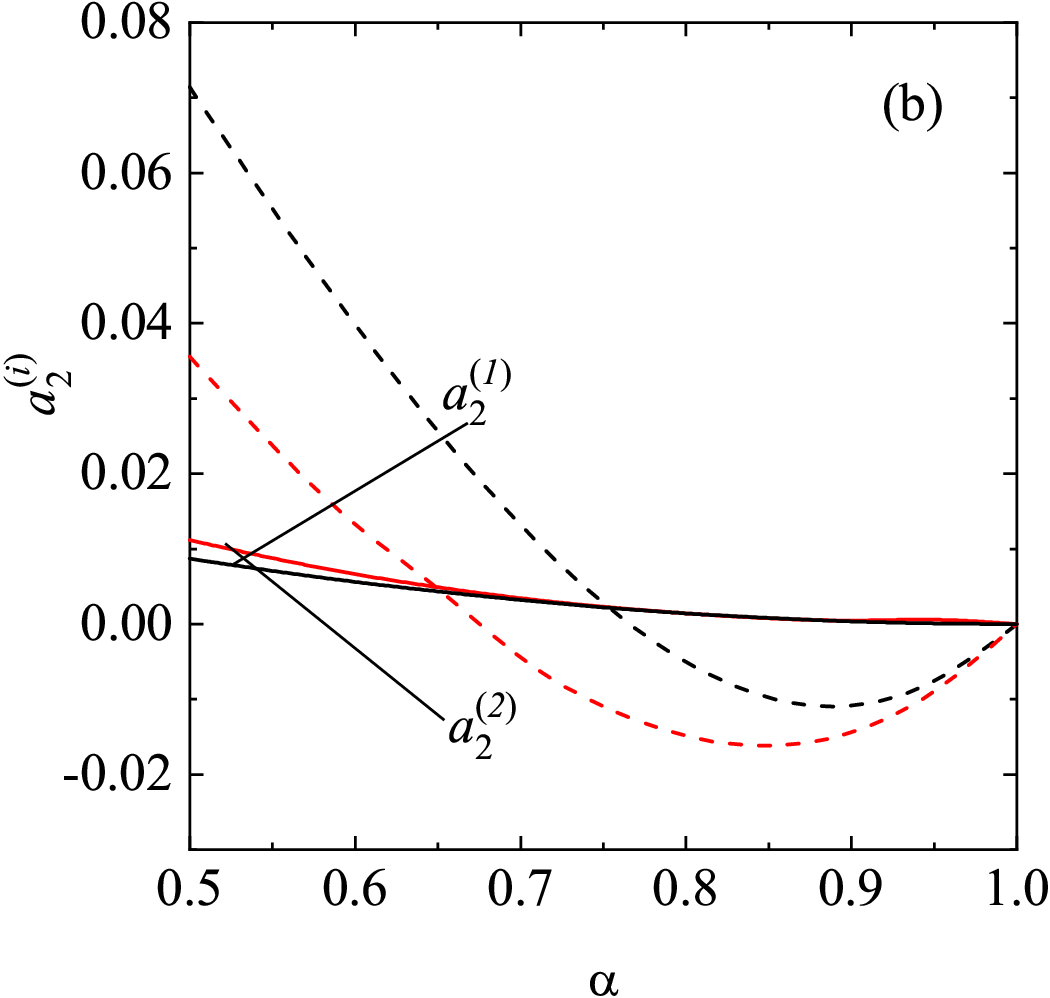}
\caption{Panel (a): Plot of the ratio $M_{4,0,0}^{*(1)}(\al)/M_{4,0,0}^{*(1)}(1)$ as a function of the (common) coefficient of restitution $\al$ for $x_1=\frac{1}{2}$, $\sigma_1/\sigma_2=1$, and $m_1/m_2=2$. The solid and dashed lines correspond to the results derived from the kinetic model and the Boltzmann equation, respectively. Panel (b): Plot of the kurtosis $a_2^{(i)}$ versus the (common) coefficient of restitution $\al$ for $x_1=\frac{1}{2}$, $\sigma_1/\sigma_2=1$, and $m_1/m_2=2$. The solid lines correspond to the results derived here from the kinetic model while the dashed lines refer to the results obtained from the Boltzmann equation.}
\label{fig5}
\end{figure}

Although the temperature ratio $T_1/T_2$ is not a hydrodynamic quantity, its dependence on the mixture's parameter space plays a crucial role in determining the transport coefficients. \cite{GD02} In fact, it is the most relevant quantity in the HCS. The temperature ratio $\gamma$ is obtained by numerically solving Eq.\ \eqref{3.7} where the partial cooling rates $\zeta_i$ are given by Eq.\ \eqref{3.6.2}. Given that this expression coincides with the one derived from the true Boltzmann equation when $f_i$ is replaced by the Maxwellian distribution \eqref{2.15.0} (with $\TT_i=T_i$ and $\mathbf{U}_i=\mathbf{0}$),
one expects that the reliability of the kinetic model for predicting $\gamma$ is quite good. To illustrate it, the temperature ratio is plotted in Fig.\ \ref{fig4} as a function of the (common) coefficient of restitution $\al_{ij}=\al$ for several mixtures. Theoretical results are compared against numerical simulation results of the Boltzmann equation \cite{MG02,CHGG22} obtained from the direct simulation Monte Carlo (DSMC) method. \cite{B94} First, an excellent agreement between theory and simulations is observed in the complete range of values of $\al$ considered. In addition, as expected the breakdown of energy equipartition is more significant as the disparity in the mass ratio increases. In general, the temperature of the heavier species is larger than that of the lighter species.

Apart from the temperature ratio, the first nonzero moments are the (dimensionless) fourth-degree moments
$M_{4,0,0}^{*(i)}=M_{0,4,0}^{*(i)}=M_{0,0,4}^{*(i)}$ and $M_{2,2,0}^{*(i)}=M_{0,2,2}^{*(i)}=M_{2,0,2}^{*(i)}$. According to Eq.\ \eqref{3.16}, $M_{4,0,0}^{*(i)}=3M_{2,2,0}^{*(i)}$. The panel (a) of Fig.\ \ref{fig5} shows the dependence of the fourth-degree moment $M_{4,0,0}^{*(1)}(\al)$ relative to its elastic value $M_{4,0,0}^{*(1)}(1)$ on the (common) coefficient of restitution $\al$ for $x_1=\frac{1}{2}$, $\sigma_1/\sigma_2=1$, and $m_1/m_2=2$. We have reduced the moment $M_{4,0,0}^{*(1)}$ with respect to its value for elastic collisions because we are mainly interested here in assessing the impact of inelasticity in collisions on the high-degree moments. For the sake of comparison, we have also plotted the corresponding result obtained from the Boltzmann equation when the scaled distribution $\varphi_i$ is approximated by its leading Sonine approximation \cite{GD99b}
\beq
\label{3.17.2}
\varphi_i(\mathbf{c})\to \pi^{-3/2}\theta_i^{3/2} e^{-\theta_i c^2}\Big[1+\frac{a_2^{(i)}}{2}\Big(\theta_i^2 c^4-5\theta_i c^2+\frac{15}{4}\Big)\Big],
\eeq
where the kurtosis $a_2^{(i)}$ is defined as \cite{GD99b}
\beq
\label{3.17.1}
a_2^{(i)}=\frac{4}{15}\theta_i^2\int  d\mathbf{c}\; c^4 \varphi_i(\mathbf{c})-1.
\eeq
The kurtosis (or fourth moment) quantifies the deviation of $\varphi_i$ from its Maxwellian form $\pi^{-3/2}\theta_i^{3/2} e^{-\theta_i c^2}$. We observe in panel (a) of Fig.\ \ref{fig5} that the prediction of the kinetic model for the fourth-degree moment $M_{4,0,0}^{*(1)}(\al)$ (relative to its value for elastic collisions) agrees quite well with that of the Boltzmann equation. In fact, the relative discrepancies between the two predictions are less than 2\% in the range of values of the coefficient of restitution considered. However, such a good agreement is not maintained when the kurtosis $a_2^{(i)}$ is considered, as shown in panel (b) of Fig.\ \ref{fig5}. The origin of this discrepancy can be partly explained by the choice of the reference function $f_{ij}$ of the kinetic model in the simpler case of a single-component granular gas. In this limiting case (where $a_2^{(1)}=a_2^{(2)}=a_2$), the kinetic model \eqref{2.22} reduces to the simplest version of the kinetic model proposed by Brey \emph{et al.}, \cite{BDS99}, where the HCS distribution function $f_\text{HCS}$ is replaced by the Maxwellian distribution with $T_1=T_2=T$. As a result of this simplification, $a_2=0$ for mechanically equivalent particles in the kinetic model for mixtures. This result can be easily obtained from the expression \eqref{3.16} for the asymptotic moments $M_{k_1,k_2,k_3}^{(i)}$ when $k_1+k_2+k_3=4$. However, as shown in the previous (approximate) results for single component granular gases derived from the Boltzmann equation for $a_2$, \cite{GS95,BRC96,NE98,MS00,BP06a,CDPT03,SM09} the magnitude of the kurtosis $a_2$ is generally very small but non-zero. Therefore, based on these results, it is expected that for granular mixtures the theoretical prediction of $a_2^{(i)}$ underestimates the value provided by the original Boltzmann equation. This trend is clearly shown in panel (b) of Fig.\ \ref{fig5}, where we observe that while the kinetic model results predict a monotonic increase of $a_2^{(i)}$ with inelasticity, the Boltzmann equation yields a non-monotonic dependence of the kurtosis on inelasticity. Furthermore, as expected, the magnitude of $a_2^{(i)}$ is much smaller in the kinetic model than in the Boltzmann equation. In any case, for practical purposes, both theoretical predictions clearly show that the coefficients $a_2^{(i)}$ are quite small and hence their impact on transport properties can be generally neglected.

\subsection{Velocity distribution function in the HCS}
\label{sec3.C}

A complete description of the HCS requires the knowledge of velocity distribution functions $f_i$. However, as said before, an explicit solution of the Boltzmann equation in the HCS is not known and the information about the distribution functions is obtained only indirectly through the (approximate) knowledge of the first few velocity moments. On the other hand, the use of a kinetic model allows in some situations to obtain the exact form of the distribution functions. Based on the good qualitative agreement found for molecular gases between the BGK results and Monte Carlo simulations, \cite{GBS90,MSG96a} one expects that the distribution functions obtained as an exact solution of the kinetic model in the HCS describes the ``true'' distributions at least in the region of thermal velocities (let's say, $\mathbf{c}\sim 1$).

The kinetic equation \eqref{3.1} for the distribution $f_1(\mathbf{v})$ in the HCS can be rewritten as
\beq
\label{3.18}
\partial_t f_1+\frac{\omega_{1}}{2}f_1-\frac{\epsilon_{1}}{2}
\frac{\partial}{\partial \mathbf{v}}\cdot \left(\mathbf{v} f_1\right)=\frac{1}{2}\Phi_1,
\eeq
where
\beq
\label{3.19.1}
\Phi_1=\omega_{11}f_{11}+\omega_{12}f_{12}.
\eeq
According to Eq.\ \eqref{3.8}, the term $\partial_t f_1$ can be expressed as
\vicente{
\beq
\label{3.19}
\frac{\partial f_1}{\partial t}=\frac{\partial f_1}{\partial T} \frac{\partial T}{\partial t}=-\zeta T\frac{\partial f_1}{\partial T},
\eeq
where $\zeta=\zeta_1=\zeta_2$ and use has been made of Eq.\ \eqref{3.5.1}. According to Eq.\ \eqref{3.8}, the dependence of the distribution $f_1$ on the granular temperature $T$ in the HCS allows us to write the identity
\beq
\label{3.19.1bis}
\frac{\partial f_1}{\partial T}=-\frac{1}{2T}\frac{\partial}{\partial \mathbf{v}}\cdot \left(\mathbf{v} f_1\right).
\eeq
Taking into account Eq.\ \eqref{3.19.1bis}, Eq. \eqref{3.19} becomes
\beq
\label{3.19.2}
\frac{\partial f_1}{\partial t}=\frac{1}{2}\zeta\frac{\partial}{\partial \mathbf{v}}\cdot \left(\mathbf{v} f_1\right),
\eeq
}
and so, Eq.\ \eqref{3.18} can be rewritten as
\beq
\label{3.20}
\Bigg(\omega_1-3\xi_1-\xi_1\mathbf{v}\cdot \frac{\partial}{\partial \mathbf{v}}\Bigg)f_1(\mathbf{v})=\Phi_1(\mathbf{v}),
\eeq
where
\beqa
\label{3.21}
\xi_1&=&\epsilon_{1}-\zeta
=\frac{8\sqrt{\pi}}{3}n_2\sigma_{12}^{2}\frac{m_1 m_2}{(m_1+m_2)^2}
(1+\al_{12})\nonumber\\
& &\times \left(\frac{2{T}_1}{m_1}+
\frac{2{T}_2}{m_2}\right)^{1/2}
\Bigg(\frac{T_2}{T_1}-1\Bigg).
\eeqa
For elastic collisions, $T_1=T_2=T_{12}=T$, $\xi_1=0$ and the solution to Eq.\ \eqref{3.20} is the Maxwellian distribution
\beq
\label{3.21.1}
f_{i,\text{el}}(\mathbf{v})=n_i\left(\frac{m_i}{2\pi T}\right)^{3/2} \exp{\left(-\frac{m_i v^2}{2T}\right)}.
\eeq

For inelastic collisions, $\xi_1\neq 0$ and the hydrodynamic (formal) solution to Eq.\ \eqref{3.21} is
\beqa
\label{3.22}
f_1(\mathbf{v})&=&\Big(\omega_1-3\xi_1-\xi_1\mathbf{v}\cdot \frac{\partial}{\partial \mathbf{v}}\Big)^{-1}\Phi_1(\mathbf{v})\nonumber\\
&=&\int_0^\infty\; ds\; e^{-(\omega_1-3\xi_1)s}e^{\xi_1 s\mathbf{v}\cdot \frac{\partial}{\partial \mathbf{v}}}\Phi_1(\mathbf{v}).
\eeqa
The action of the scaling operator $e^{a\mathbf{v}\cdot \frac{\partial}{\partial \mathbf{v}}}$ on an arbitrary function $F(\mathbf{v})$ is
\beq
\label{3.23}
e^{a\mathbf{v}\cdot \frac{\partial}{\partial \mathbf{v}}}F(\mathbf{v})=F(e^{a} \mathbf{v}).
\eeq
\begin{widetext}
Equation \eqref{3.22} can be more explicitly written when one takes into account the relationship \eqref{3.23}:
\vicente{
\beqa
\label{3.24}
f_1(\mathbf{v})&=&\int_0^\infty\; ds\; e^{-(\omega_1-3\xi_1)s} \Phi_1\left(e^{\xi_1 s}\mathbf{v}\right)\nonumber\\
&=&
\int_0^\infty\; ds\; e^{-(\omega_1-3\xi_1)s} n_1\Bigg[\omega_{11}\Big(\frac{m_1}{2T_1}\Big)^{3/2}\exp\Big(-\frac{m_1}{2T_1}e^{2\xi_1 s}v^2\Big)+\omega_{12}\Big(\frac{m_1}{2T_{12}}\Big)^{3/2}\exp\Big(-\frac{m_1}{2T_{12}}e^{2\xi_1 s}v^2\Big)\Bigg].
\eeqa
Equation \eqref{3.24} can be expressed in dimensionless form by introducing the dimensionless time $\tau=\nu s$. In
terms of the dimensionless quantities $\omega_1^*=\omega_1/\nu$,  $\xi_1^*=\xi_1/\nu$, and $\mathbf{c}=\mathbf{v}/v_\text{th}$, one writes $f_1$ in the form \eqref{3.8} where the scaled distribution $\varphi_1(\mathbf{c})$ is given by}
\beq
\label{3.25}
\varphi_1(\mathbf{c})=\pi^{-3/2}\int_0^\infty\; d\tau\; e^{-(\omega_1^*-3\xi_1^*)\tau}\Big[\omega_{11}^*\theta_1^{3/2}
\exp\Big(-\theta_1 e^{2\xi_1^*\tau}c^2\Big)+\omega_{12}^*\theta_{12}^{3/2}
\exp\Big(-\theta_{12} e^{2\xi_1^*\tau}c^2\Big)\Big].
\eeq
\end{widetext}
The corresponding expression for $\varphi_2$ can be easily obtained by making the change $1\leftrightarrow 2$.

The knowledge of the scaled distribution $\varphi_1$ allows us to compute the dimensionless moments $M_{k_1,k_2,k_3}^{*(1)}$ defined as
\beq
\label{3.25.1}
M_{k_1,k_2,k_3}^{*(1)}=\int d\mathbf{c}\; c_x^{k_1}c_y^{k_2}c_z^{k_3}\varphi_1(\mathbf{c}).
\eeq
Further technical details of this evaluation are provided in the Appendix \ref{appA}. As anticipated, the corresponding expression for $M_{k_1,k_2,k_3}^{*(1)}$ is consistent with Eq.\ \eqref{3.16}, confirming the coherence of the results presented here for the HCS

According to Eq.\ \eqref{3.25}, we observe that $\varphi_1$ diverges to infinity at $\mathbf{c}=\mathbf{0}$ when $\omega_1^*\leq 3\xi^*$. This singularity primarily arises from the collisional dissipation due to the inelastic nature of the collisions. As seen in Eq.\ \eqref{3.25}, two competing exponential terms appear in the form of the distribution $\varphi_1$. The term $e^{-\omega_1^*\tau}$ essentially represents the fraction of particles of species 1 that have not collided after $\tau$ effective collision times, while $e^{3\xi_1^*\tau}$ results from the inelasticity of the collisions. In the quasielastic limit ($\alpha_{ij}\lesssim 1$, where $\omega_1^*>3\xi_1^*$), the collisional dissipation is not large enough to dominate the effects of the collisions, and thus $\varphi_1$ remains finite at $\mathbf{c}=\mathbf{0}$. However, if the inelasticity is strong enough that $3\xi_1^*\geq \omega_1^*$, the opposite occurs, leading to a ``condensation'' of particles of species 1 around $\mathbf{c}=\mathbf{0}$.

To illustrate the dependence of $\varphi_1(\mathbf{c})$ on the (dimensionless) velocity $\mathbf{c}$, let us consider the marginal distribution
\begin{widetext}
\beqa
\label{3.26}
\varphi_{1,x}(c_x)&=&\int_{-\infty}^{+\infty}dc_y\int_{-\infty}^{+\infty}dc_z\; \varphi_1(\mathbf{c})\nonumber\\
&=&\pi^{-1/2}\int_0^\infty d\tau\; e^{-(\omega_1^*-\xi_1^*)\tau}\Big[\omega_{11}^*\theta_1^{1/2}
\exp\Big(-\theta_1 e^{2\xi_1^*\tau}c_x^2\Big)+\omega_{12}^*\theta_{12}^{1/2}
\exp\Big(-\theta_{12} e^{2\xi_1^*\tau}c_x^2\Big)\Big].
\eeqa
\end{widetext}
For elastic collisions ($\al_{ij}=1$), $\xi_1^*=0$, $\theta_1=\theta_{12}=2\mu_{12}$, and Eq.\ \eqref{3.26} becomes
\beq
\label{3.27}
\varphi_{1,x}^\text{el}(c_x)=\pi^{-1/2}(2\mu_{12})^{1/2}e^{-2\mu_{12}c_x^2}.
\eeq
\begin{figure}[ht]
\includegraphics[width=0.4\textwidth]{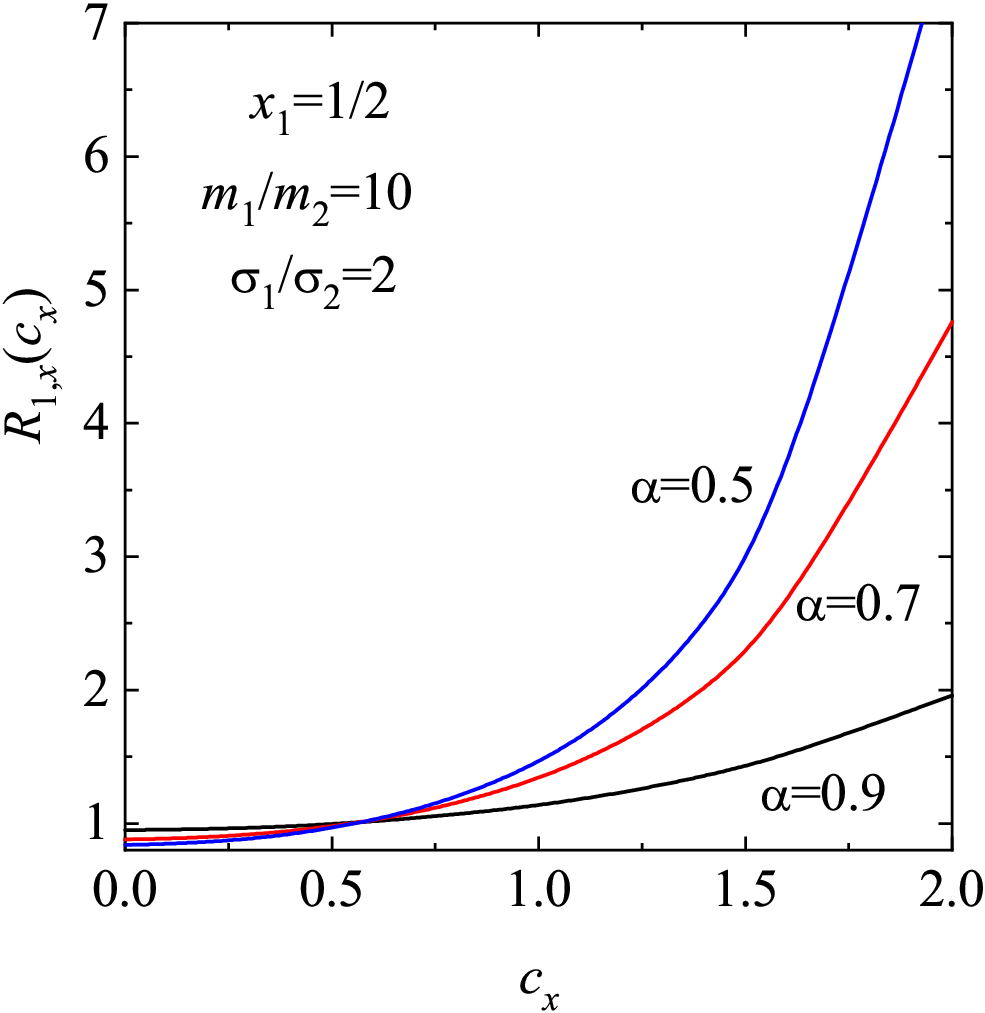}
\caption{Plot of the ratio $R_{1,x}(c_x)=\varphi_{1,x}(c_x)/\varphi_{1,x}^\text{el}(c_x)$ versus the (scaled) velocity $c_x$ for $x_1=\frac{1}{2}$, $m_1/m_2=10$, $\sigma_1/\sigma_2=2$, and three different values of the (common) coefficient of restitution $\al_{ij}=\al$: $\al=0.9$, 0.7 and 0.5.}
\label{fig6}
\end{figure}

Figure \ref{fig6} plots the ratio $R_{1,x}(c_x)=\varphi_{1,x}(c_x)/\varphi_{1,x}^\text{el}(c_x)$ as a function of the (scaled) velocity $c_x$ for $x_1=\frac{1}{2}$, $m_1/m_2=10$, $\sigma_1/\sigma_2=2$, and three different values of the (common) coefficient of restitution $\al_{ij}=\al$. In the cases considered in Fig.\ \ref{fig6}, $\omega_1^*-\xi_1^*>0$ and hence $\varphi_{1,x}(c_x)$ remains finite at $c_x=0$.
As expected, we observe that the deviation from the Maxwellian distribution function ($R_{1,x} = 1$) becomes more pronounced as inelasticity increases. Additionally, for sufficiently large velocities, the population of particles relative to its elastic value increases as the coefficient of restitution decreases.

\section{Chapman--Enskog method. Navier--Stokes transport coefficients}
\label{sec4}

Once the HCS is well characterized, the next step is to determine the Navier--Stokes transport coefficients of the mixture. These coefficients can be obtained by solving the kinetic model  \eqref{2.22} by means of the application of the Chapman-Enskog method \cite{CC70} conveniently generalized to inelastic collisions. Since the extension of this method to granular gases has been extensively discussed in some previous works (see for example, Ref.\ \onlinecite{G19}), only some details on its application to granular mixtures are given in the Appendix \ref{appB}.

It is quite obvious that the balance equations \eqref{2.9}--\eqref{2.11} become a closed set of hydrodynamic equations for the fields $n_i$, $\mathbf{U}$ and $T$ once the mass, momentum and heat fluxes [defined by Eqs.\ \eqref{2.12}--\eqref{2.14}, respectively] and the cooling rate $\zeta$ [defined by Eq.\ \eqref{2.8}] are expressed in terms of the hydrodynamic fields and their gradients. As discussed in previous works, \cite{GD02,GMD06} while the pressure tensor has the same form as for a one-component system, there is greater freedom in the representation of the heat and mass fluxes. Here, as in Refs.\ \onlinecite{GD02,GMD06,GM07}, we take the gradients of the mole fraction $x_1=n_1/n$,
the pressure $p=nT$, the temperature $T$, and the flow velocity ${\bf U}$ as the relevant hydrodynamic fields.

For times longer than the mean free time (where the granular gas mixture has completely ``forgotten'' the details of its initial preparation) and for regions far from the boundaries of the system, the granular mixture is expected to reach a hydrodynamic regime. In this regime, the Boltzmann kinetic equation admits a special solution, called the \emph{normal} (or hydrodynamic) solution, characterized by the fact that the distribution functions $f_i$ depend on space and time only through a functional dependence on the hydrodynamic fields $\varpi\equiv \left(x_1, \mathbf{U}, p, T\right)$. For simplicity, this functional dependence can be made local in space and time when the spatial gradients of $\varpi$ are small and one can write $f_i$ as a series expansion in a formal parameter $\varepsilon$, which measures the nonuniformity of the system:
\begin{equation}
f_{i}=f_{i}^{(0)}+\varepsilon \,f_{i}^{(1)}+\varepsilon^2 \,f_{i}^{(2)}+\cdots \;,
\label{3.31}
\end{equation}
where each factor of $\varepsilon $ means an implicit gradient of a hydrodynamic field. The local reference state $f_i^{(0)}$ is chosen to give the same first moments as the exact distribution $f_i$.

Use of the Chapman--Enskog expansion (\ref{3.31}) in the definitions of the fluxes (\ref{2.12})--(\ref{2.14}) and the cooling rate (\ref{2.7}) gives the corresponding expansion for these quantities.  The time derivatives
of the fields are also expanded as $\partial_t=\partial_t^{(0)}+\epsilon
\partial_t^{(1)}+\cdots$. The coefficients of the time derivative expansion are identified from the balance equations (\ref{2.9})--(\ref{2.11}) after
expanding the fluxes and the cooling rate $\zeta$. This is the usual procedure of the Chapman--Enskog method. \cite{CC70}

\subsection{Zeroth-order distribution function}

In the zeroth order, $f_i^{(0)}$ obeys the kinetic equation
\beq
\label{3.32}
\partial_t^{(0)} f_i^{(0)}+\frac{1}{2}\omega_i f_i^{(0)}-\frac{\epsilon_{i}}{2}\frac{\partial}{\partial \mathbf{V}}\cdot \mathbf{V}f_i^{(0)}=\frac{1}{2}\sum_{j=1}^2\omega_{ij}f_{ij}^{(0)},
\eeq
where
\beq
\label{3.33}
f_{ij}^{(0)}(\mathbf{V})=n_i\left(\frac{m_i}{2\pi
T_{i}}\right)^{3/2} \exp\left(-\frac{m_i}{2T_{i}}V^2\right).
\eeq

The macroscopic balance equations to zeroth-order give
\beq
\label{3.34}
\partial_t^{(0)}x_1=0, \quad \partial_t^{(0)}\mathbf{U}=\mathbf{0}, \quad \partial_t^{(0)}\ln T=\partial_t^{(0)}\ln p=-\zeta^{(0)},
\eeq
where $\zeta^{(0)}$ is the zeroth-order approximation to the cooling rate. Its form is given by Eq.\ \eqref{3.6.1}. Since $f_i^{(0)}$ is a normal solution, according to Eq.\ \eqref{3.34}, then
\beq
\label{3.35}
\partial_t^{(0)} f_i^{(0)}=\partial_Tf_i^{(0)}\partial_t^{(0)}T+\partial_pf_i^{(0)}\partial_t^{(0)}p=\frac{1}{2}\zeta^{(0)}\frac{\partial}{\partial \mathbf{V}}\cdot \left(\mathbf{V}f_i^{(0)}\right),
\eeq
where $f_i^{(0)}$ has been assumed to be of the form \eqref{3.8}. Substitution of Eq.\ \eqref{3.35} into Eq.\ \eqref{3.32} yields
\beq
\label{3.36}
\frac{1}{2}\zeta^{(0)}\frac{\partial}{\partial \mathbf{V}}\cdot \left(\mathbf{V}f_i^{(0)}\right)+\frac{1}{2}\omega_i f_i^{(0)}-\frac{\epsilon_{i}}{2}\frac{\partial}{\partial \mathbf{V}}\cdot \mathbf{V}f_i^{(0)}=\frac{1}{2}\sum_{j=1}^2\omega_{ij}f_{ij}^{(0)}.
\eeq
Equation \eqref{3.36} has the same form as that of the HCS [see Eqs.\ \eqref{3.18} and \eqref{3.19.2}] except that $f_i^{(0)}(\mathbf{r}, \mathbf{v};t)$ is the \emph{local} HCS distribution function of the species $i$. Thus, the distribution function $f_i^{(0)}$ is given by Eq.\ \eqref{3.8} with the replacements $n_i\to n_i(\mathbf{r}, t)$, $\mathbf{v}\to \mathbf{V}=\mathbf{v}-\mathbf{U}(\mathbf{r}, t)$, and $T\to T(\mathbf{r}, t)$. Since $f_i^{(0)}$ is isotropic in $\mathbf{V}$, it follows that
\begin{equation}
\label{3.37}
{\bf j}_1^{(0)}={\bf 0},\quad{\bf q}^{(0)}={\bf 0},\quad P_{ij}^{(0)}=p\delta_{ij},
\end{equation}
where we recall that $p=nT$.

\subsection{Transport coefficients}

The derivation of the kinetic equation obeying the first-order distributions is quite large and follows similar mathematical steps as those previously made in the original Boltzmann equation. \cite{GD02,GM07} Some specific details on this calculation as well as on the determination of the Navier--Stokes transport coefficients are offered in the Appendix \ref{appB}. For the sake of brevity, only the final expressions are displayed in this section.

To first order in the spatial gradients, the phenomenological constitutive
relations for the fluxes in the low-density regime have the forms \cite{GM84}
\begin{equation}
\label{3.28}
{\bf j}_1^{(1)}=-\frac{m_1m_2n}{\rho}D\nabla x_1-\frac{\rho}{p}D_p\nabla p-
\frac{\rho}{T}D_T\nabla T,
\eeq
\begin{equation}
\label{3.29}
P_{k\ell}^{(1)}=p\delta_{k\ell}-\eta\left(\nabla_k U_\ell+
\nabla_\ell U_k-\frac{2}{d}\delta_{k\ell}\nabla \cdot {\bf U}\right),
\end{equation}
\begin{equation}
\label{3.30}
{\bf q}^{(1)}=-T^2D''\nabla x_1-L_p\nabla p-\lambda_T\nabla T.
\end{equation}
In Eqs.\ \eqref{3.28}--\eqref{3.30}, the transport coefficients are the diffusion coefficient $D$, the thermal
diffusion coefficient $D_T$, the pressure diffusion coefficient $D_p$, the shear viscosity $\eta$, the
Dufour coefficient $D''$, the thermal conductivity $\lambda_T$, and the pressure
energy coefficient $L_p$.

\subsubsection{Diffusion transport coefficients}

The expressions of the transport coefficients associated with the mass flux are
\beqa
\label{3.38}
D&=&\frac{\rho}{m_{1}m_{2}n}\left(\nu_D -\frac{1}{2}\zeta ^{(0)}\right) ^{-1}
\Bigg[ p\left(\frac{\partial}{\partial x_{1}}x_{1}\gamma_{1}\right)_{p,T}
\nonumber\\
& &
+\rho
\left(\frac{\partial \zeta ^{(0)}}{\partial x_{1}}\right)_{p,T}
\left(
D_{p}+D_T\right) \Bigg] ,
\eeqa
\begin{equation}
\label{3.39}
D_{p}=\frac{n_{1}T_{1}}{\rho}\left(1-\frac{m_{1}nT}{\rho T_{1}}\right)
\left(\nu_D -\frac{3}{2}\zeta^{(0)}+\frac{\zeta^{(0)2}}{2\nu_D}\right)^{-1},
\end{equation}
\begin{equation}
\label{3.40}
D_T=-\frac{\zeta^{(0)}}{2\nu }D_{p}.
\end{equation}
In Eqs.\ \eqref{3.38} and  \eqref{3.39} we have introduced the collision frequency
\beqa
\label{3.40.1}
\nu_D&=&\frac{1}{2}\left(\mu_{21}+\frac{x_1}{x_2}\mu_{12}\right)\omega_{12}
\nonumber\\
&=&
\frac{4\sqrt{\pi}}{3}\frac{\rho}{n(m_1+m_2)}
\left(\frac{\theta_1+\theta_2}{\theta_1\theta_2}\right)^{1/2}(1+\alpha_{12})\nu.\nonumber\\
\eeqa
It must be remarked that the expressions \eqref{3.38}--\eqref{3.40} are identical to those obtained from the Boltzmann equation in the first-Sonine approximation when one neglects non-Gaussian corrections to the distributions $f_i^{(0)}$ (i.e., $a_2^{(i)}=0)$. \cite{GD02,GM07} This agreement is in fact a consequence of one of the requirements of the kinetic model. The coefficient $D$ is symmetric while the coefficients $D_p$ and $D_T$ are antisymmetric under the change $1\leftrightarrow 2$. As a consequence of these symmetries, ${\bf j}_2^{(1)}=-{\bf j}_1^{(1)}$ as expected. For mechanically equivalent particles ($m_1=m_2, \sigma_1=\sigma_2, \al_{ij}=\al$), $\gamma_i=\gamma=1$, $D_p=D_T=0$, and
\beq
\label{3.41}
D_{\text{self}}=\frac{3}{2}\frac{1}{\sigma^2}\sqrt{\frac{T}{\pi m}}
\frac{1}{(1+\al)^2}
\eeq
is the self-diffusion coefficient.

\subsubsection{Shear viscosity coefficient}

The shear viscosity coefficient $\eta$ can be written as
\beq
\label{3.42}
\eta=p\Big(\frac{x_1\gamma_1}{\beta_1-\frac{1}{2}\zeta^{(0)}}+\frac{x_2\gamma_2}{\beta_2-\frac{1}{2}\zeta^{(0)}}\Big),
\eeq
where
\beq
\label{3.42.1}
\beta_i=\frac{\omega_i+\epsilon_i}{2}.
\eeq
For mechanically equivalent particles, Eq.\ \eqref{3.42} leads to the expression given by the model for the shear viscosity of a dilute granular gas:
\beq
\label{3.43}
\eta=\frac{p}{\beta-\frac{1}{2}\zeta^{(0)}},
\eeq
where
\beq
\label{3.43.1}
\beta=\frac{\sqrt{2\pi}}{3}(5-\al)(1+\al)\nu, \quad
\zeta^{(0)}=\frac{2\sqrt{2\pi}}{3}(1-\al^2)\nu.
\eeq
The expression \eqref{3.43} matches the one derived from the original Boltzmann equation \cite{BDKS98} \vicente{in the \emph{standard} first Sonine approximation when $\beta$ is replaced by the collision frequency
\beq
\label{3.43.2}
\beta_{\eta}^{\text{BE}}=\frac{2}{5}\sqrt{2\pi}(3-\al)(1+\al)\Big(1+\frac{7}{16}a_2\Big)\nu,
\eeq
where the kurtosis $a_2$ is \cite{NE98}
\beq
\label{3.43.3}
a_2=\frac{16(1-\al)(1-2\al^2)}{81-17\al+30(1-\al)\al^2}.
\eeq
}

\subsubsection{Heat flux transport coefficients}

As usual, the study of the heat flux is much more involved. Its constitutive equation is given by Eq.\ \eqref{3.30} where the transport coefficients are
\beq
\label{3.44}
D''=D_1''+D_2'', \quad L_p=L_{p,1}+L_{p,2}, \quad \lambda_T=\lambda_{T,1}+\lambda_{T,2}.
\eeq
Here, in contrast to the results derived from the Boltzmann equation, \cite{GD02,GM07} the equation defining the partial contributions of the species $1$ ($D_1''$, $L_{p,1}$, and $\lambda_{T,1}$) are decoupled from their corresponding counterparts of the species $2$. In the case of the species 1 and by using matrix notation, the coupled set of three equations for the unknowns
\beq
\label{3.45}
\left\{D_1'',L_{p,1}, \lambda_{T,1} \right\}
\eeq
can be written as
\begin{equation}
\label{3.46}
\Lambda_{1,\sigma \sigma'}X_{1,\sigma'}=Y_{1,\sigma}.
\end{equation}
Here, $X_{1,\sigma'}$ is the column matrix defined by the set \eqref{3.45}, $\Lambda_{1,\sigma\sigma'}$ is the square matrix
\begin{equation}
\label{3.47}
\boldsymbol{\Lambda}_1=\left(
\begin{array} {ccc}
T^2(\frac{3}{2}\zeta^{(0)}-\beta_{1})&
p\left(\frac{\partial \zeta ^{(0)}}{\partial x_{1}}\right)_{p,T}&
T\left( \frac{\partial \zeta ^{(0)}}{\partial x_{1}}\right)_{p,T} \\
0& \frac{5}{2}\zeta^{(0)}-\beta_{1}&
T\zeta^{(0)}/{p}\\
0& -p\zeta^{(0)}/2T&\zeta^{(0)}-\beta_{1}
\end{array}
\right),
\end{equation}
and the column matrix $\boldsymbol{Y}_1$ is
\begin{equation}
\label{3.48}
\boldsymbol{Y}_1=\left(
\begin{array}{c}
-\frac{5}{4}\frac{m_1m_2n}{\rho}A_{1}D-
\frac{5}{2}\frac{nT^2}{m_1}\frac{\partial}{\partial x_1}\left(x_1\gamma_1^2\right)\\
-\frac{5}{4}\frac{\rho}{p}A_{1}D_p-
\frac{5}{2}
\frac{n_1T_1^2}{m_1p}
\left(1-\frac{m_1p}{\rho T_1}\right)\\
-\frac{5}{4}\frac{\rho}{T}A_{1}D_T-
\frac{5}{2}
\frac{n_1T_1^2}{m_1T}
\end{array}
\right).
\end{equation}
In Eq.\ \eqref{3.48},
\beq
\label{3.49}
A_1=\omega_{11}\frac{T_1}{m_1}+\omega_{12}\mu_{12}\frac{n_2-n_1}{n_2}\frac{T_{12}}{m_1}+\epsilon_{1}\frac{T_1}{m_1}.
\eeq
Analogously, the matrix equation defining the unknowns
\beq
\label{3.50}
\left\{D_2'',L_{p,2}, \lambda_{T,2} \right\}
\eeq
can be written
\begin{equation}
\label{3.51}
\Lambda_{2,\sigma \sigma'}X_{2,\sigma'}=Y_{2,\sigma},
\end{equation}
where $X_{2,\sigma'}$ is the column matrix defined by the set \eqref{3.50}, $\Lambda_{2,\sigma\sigma'}$ is the square matrix
\begin{equation}
\label{3.52}
\boldsymbol{\Lambda}_2=\left(
\begin{array} {ccc}
T^2(\frac{3}{2}\zeta^{(0)}-\beta_{2})&
p\left(\frac{\partial \zeta ^{(0)}}{\partial x_{1}}\right)_{p,T}&
T\left( \frac{\partial \zeta ^{(0)}}{\partial x_{1}}\right)_{p,T} \\
0& \frac{5}{2}\zeta^{(0)}-\beta_{2}&
T\zeta^{(0)}/{p}\\
0& -p\zeta^{(0)}/2T&\zeta^{(0)}-\beta_{2}
\end{array}
\right),
\end{equation}
and the column matrix $\boldsymbol{Y}_2$ is
\begin{equation}
\label{3.53}
\boldsymbol{Y}_2=\left(
\begin{array}{c}
\frac{5}{4}\frac{m_1m_2n}{\rho}A_{2}D-
\frac{5}{2}\frac{nT^2}{m_1}\frac{\partial}{\partial x_1}\left(x_2\gamma_2^2\right)\\
\frac{5}{4}\frac{\rho}{p}A_{2}D_p-
\frac{5}{2}
\frac{n_2T_2^2}{m_2p}
\left(1-\frac{m_2p}{\rho T_2}\right)\\
\frac{5}{4}\frac{\rho}{T}A_{2}D_T-
\frac{5}{2}
\frac{n_1T_1^2}{m_1T}
\end{array}
\right).
\end{equation}
Here, $A_2$ is given from Eq.\ \eqref{3.49} by making the change $1\leftrightarrow 2$.

As expected, Eqs.\
\eqref{3.46}--\eqref{3.53} show that the Dufour coefficient $D''$ is antisymmetric with respect to the change $1\leftrightarrow 2$ while the coefficients $L_p$ and $\lambda_T$ are symmetric. The first property implies necessarily that $D''$ vanishes for mechanically equivalent particles. In this limiting case, Eqs.\
\eqref{3.46}--\eqref{3.53} yield the following expression for the heat flux:
\beq
\label{3.54}
\mathbf{q}^{(1)}=-\kappa \nabla T-\mu \nabla n,
\eeq
where
\beq
\label{3.55}
\kappa=\lambda_T+n L_p=\frac{5}{2}\frac{p}{m}\frac{1}{\beta-2\zeta^{(0)}},
\eeq
\beq
\label{3.56}
\mu=T n L_p=\frac{T}{n}\frac{\zeta^{(0)}\kappa}{\beta-\frac{3}{2}\zeta^{(0)}}.
\eeq
Upon writing Eq.\ \eqref{3.54} use has been made of the relation $\nabla p=n\nabla T+T\nabla n$.
As in the case of shear viscosity, the expressions \eqref{3.55} and \eqref{3.56} for $\kappa$ and $\mu$ are consistent with those obtained from the Boltzmann equation \cite{BDKS98} in the standard first Sonine approximation when \vicente{one takes $a_2 = 0$ and $\beta$ is replaced by the collision frequency
\beq
\label{3.56.1}
\beta_{\kappa}^{\text{BE}}=\frac{11}{30}\sqrt{2\pi}(1+\al)\Big(\frac{49}{33}-\al+\frac{19-3\al}{1056}a_2\Big)\nu,
\eeq
where $a_2$ is defined by Eq.\ \eqref{3.43.3}.} Moreover, in contrast to the results obtained from IMM, \cite{S03,BGM10} the heat flux transport coefficients are well defined functions (i.e., they are always positive) for monocomponent granular gases in the complete range of values of the coefficient of restitution $\al$.

\subsection{Comparison with the transport coefficients of the Boltzmann equation}

\begin{figure}[h]
\includegraphics[width=0.4\textwidth]{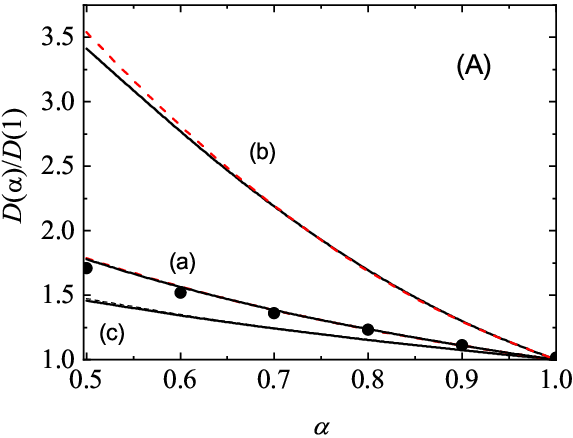}
\includegraphics[width=0.4 \textwidth,angle=0]{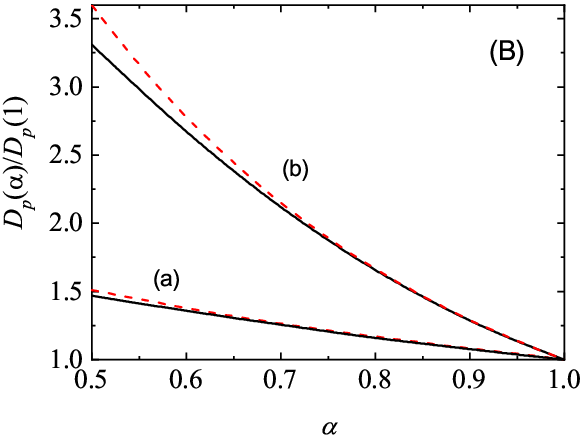}
\caption{Panel (A): Plot of the (reduced) diffusion coefficient $D(\al)/D(1)$ as a function of the (common) coefficient of restitution $\al$ for three different mixtures: $m_1/m_2=1$, $\sigma_1/\sigma_2=1$ (a), $x_1=0.2$, $m_1/m_2=4$, $\sigma_1/\sigma_2=1$ (b), and $x_1=0.2$, $m_1/m_2=0.5$, $\sigma_1/\sigma_2=1$ (c). Here, $D(1)$ is the value of the diffusion coefficient for elastic collisions. The solid lines refer to the results obtained here from the kinetic model while the dashed lines are the Boltzmann results obtained in the first Sonine approximation. \cite{GD02,GM07} The symbols refer to Monte Carlo simulations. \cite{BRCG00,GM04} Panel (B): Plot of the (reduced) pressure diffusion coefficient $D_p(\al)/D_p(1)$ as a function of the (common) coefficient of restitution $\al$ for two different mixtures:  $x_1=0.2$, $m_1/m_2=4$, $\sigma_1/\sigma_2=1$ (a), and $x_1=0.2$, $m_1/m_2=0.5$, $\sigma_1/\sigma_2=1$ (b). Here, $D_p(1)$ is the value of the pressure diffusion coefficient for elastic collisions. The solid lines refer to the results obtained here from the kinetic model while the dashed lines are the Boltzmann results obtained in the first Sonine approximation. \cite{GD02,GM07}}
\label{fig7}
\end{figure}

Let us compare the predictions of the kinetic model for the Navier-Stokes transport coefficients with those derived from the original inelastic Boltzmann equation. \cite{GD02,GMD06,GM07} We first consider the three diffusion coefficients $D$, $D_p$ and $D_T$. As mentioned before, the differences between the two descriptions for these coefficients are only due to the non-zero values of the kurtosis $a_2^{(i)}$. Since the magnitude of these coefficients is generally small (except for rather extreme values of dissipation), one expects a very good agreement between the kinetic model and the Boltzmann equation for the diffusion transport coefficients. This is illustrated in panels (A) and (B) of Fig.\ \ref{fig7} for the reduced diffusion coefficients $D(\al)/D(1)$ and $D_p(\al)/D_p(1)$; an excellent agreement between the two theories is observed except for very strong inelasticities. For example, at $\al=0.5$ the discrepancies for $D(\al)/D(1)$ and $D_p(\al)/D_p(1)$ are about $4\%$ and $5\%$, respectively, for $m_1/m_2=4$, while they are about $1\%$ and $3\%$, respectively, for $m_1/m_2=0.5$.

\begin{figure}[h]
\includegraphics[width=0.4\textwidth]{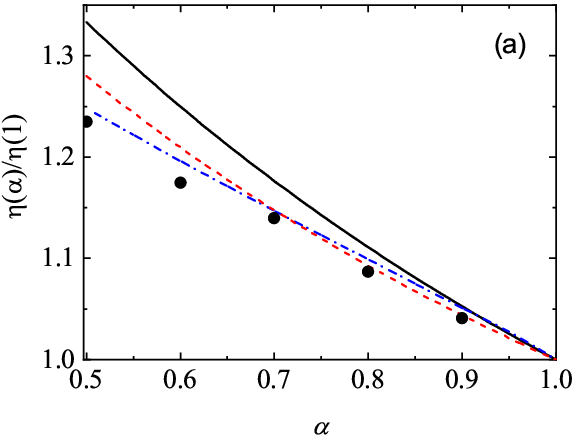}
\includegraphics[width=0.4 \textwidth,angle=0]{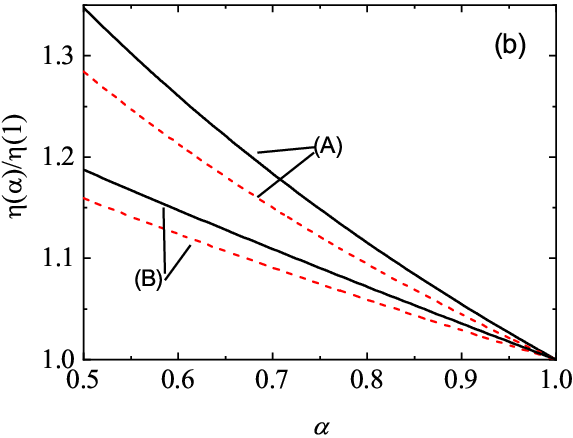}
\caption{Panel (a): Plot of the (reduced) shear viscosity coefficient $\eta(\al)/\eta(1)$ as a function of the coefficient of restitution $\al$ for a monocomponent granular gas. Here, $\eta(1)$ refers to the shear viscosity coefficient for elastic collisions. The solid and dashed lines correspond to the results derived from the kinetic model and the Boltzmann equation in the standard first Sonine approximation. \cite{BDKS98} The dash-dotted line refers to the results obtained from the Boltzmann equation by using the modified first Sonine approximation. \cite{GSM07} The symbols correspond to Monte Carlo simulations. \cite{MSG07} Panel (b): Plot of the (reduced) shear viscosity coefficient $\eta(\al)/\eta(1)$ as a function of the coefficient of restitution $\al$ for two different mixtures: $x_1=\frac{1}{2}$, $m_1/m_2=2$, $\sigma_1/\sigma_2=2$, $\al_{ij}=\al$ (A), and $x_1=\frac{1}{2}$, $m_1/m_2=2$, $\sigma_1/\sigma_2=2$, $\al_{11}=\al$, $\al_{12}=(1+\al)/2$, and $\al_{22}=(3+\al)/4$ (B). The solid lines refer to the results obtained here from the kinetic model while the dashed lines are the Boltzmann results obtained in the first Sonine approximation. \cite{GD02,GM07}}
\label{fig8}
\end{figure}
\begin{figure}[h]
\includegraphics[width=0.4\textwidth]{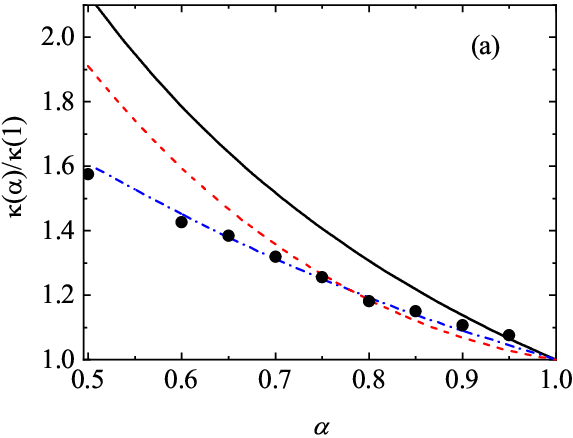}
\includegraphics[width=0.4 \textwidth,angle=0]{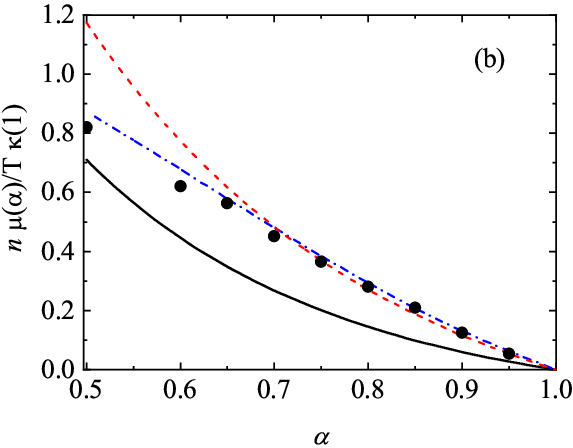}
\caption{Panel (a): Plot of the (reduced) thermal conductivity coefficient $\kappa(\al)/\kappa(1)$ as a function of the coefficient of restitution $\al$ for a monocomponent granular gas. Here, $\kappa(1)$ refers to the thermal conductivity coefficient for elastic collisions. The solid and dashed lines correspond to the results derived from the kinetic model and the Boltzmann equation in the standard first Sonine approximation. \cite{BDKS98} The dash-dotted line refers to the results obtained from the Boltzmann equation by using the modified first Sonine approximation. \cite{GSM07} The symbols correspond to Monte Carlo simulations. \cite{BRMG05} Panel (b): The same as in the panel (a) but for the (reduced) diffusive heat conductivity coefficient $n\mu(\al)/T\kappa(1)$.
}
\label{fig9}
\end{figure}

We now consider the shear viscosity coefficient $\eta$. Figure \ref{fig8} shows the (reduced) coefficient $\eta(\al)/\eta(1)$ for a single component granular gas (panel (a)) and two different mixtures (panel (b)). In the cases studied here, although the kinetic model tends to overestimate the Boltzmann results, the agreement between the two approaches is reasonably good for moderate dissipation values (e.g., $\al \gtrsim 0.8$). As expected, the relative differences increase with increasing dissipation. Moreover, the combined effect of mass and diameter ratios on these differences shows very little sensitivity, indicating that the model captures the influence of both $m_1/m_2$ and $\sigma_1/\sigma_2$ on the shear viscosity quite well. It is also worth noting that for the single component granular gas (panel (a)), the so-called modified first Sonine approximation (an approximation where the Maxwellian distribution for the zeroth-order approximation $f^{(0)}$ is replaced by the HCS distribution) \cite{GSM07} shows better agreement with the simulation data than the standard first Sonine approach. \cite{GD02}

More significant discrepancies between the kinetic model and the Boltzmann equation are expected at the level of the heat flux transport coefficients. To illustrate this, for the sake of simplicity, we consider the single component granular gas. Figure \ref{fig9} shows the $\al$ dependence of the (reduced) heat flux transport coefficients $\kappa(\al)/\kappa(1)$ and $n\mu(\al)/T\kappa(1)$. The coefficients $\kappa$ and $\mu$ are defined by Eqs.\ \eqref{3.55} and \eqref{3.56}, respectively.
In the case of elastic collisions ($\al=1$) the diffusive heat conductivity coefficient $\mu$ vanishes. We find that the kinetic model qualitatively captures the trends observed in the original Boltzmann equation. On a more quantitative level, however, the discrepancies become more pronounced: while the model overestimates the Boltzmann values of $\kappa$, it underestimates the Boltzmann values of $\mu$. Figure \ref{fig9} also shows the disagreement between the standard first Sonine approximation and computer simulations for cases with strong dissipation. These differences are significantly reduced by the modified first Sonine approximation. \cite{GSM07}

In summary, the kinetic model captures, at least on a semi-quantitative level, the influence of inelasticity on the Navier--Stokes transport coefficients. In particular, the three diffusion coefficients ($D$, $D_p$ and $D_T$) are almost the same in both Boltzmann and model kinetic equations, while the shear viscosity $\eta$ is underestimated by the kinetic model. More pronounced discrepancies occur in the case of the heat flux transport coefficients. For example, in the limiting case of a one-component granular gas, the relative differences between the kinetic model and the original Boltzmann equation for the thermal conductivity $\kappa$ are about 11\% at $\al=0.5$ and 10\% at $\al=0.8$.

\subsection{Origin of the discrepancies between the kinetic model and Boltzmann results}

\vicente{
In an attempt to understand the origin of the discrepancies observed, particularly in the case of the shear viscosity and thermal conductivity coefficients, let us consider the limiting case of a simple granular gas ($m_1=m_2$, $\sigma_1=\sigma_2$, and $\al_{11}=\al_{22}=\al_{12}$).

In this limiting case, as mentioned in previous works, \cite{GS03} one of the main drawbacks of the kinetic model is that all relaxation processes are accounted for by only a single relaxation time. In other words, in the kinetic model, all non-zero eigenvalues of the linearized Boltzmann collision operator are collapsed into a single eigenvalue. This implies that the shear viscosity coefficient $\eta$, the heat conductivity coefficient $\kappa$, and the diffusive heat conductivity coefficient $\mu$ are given in terms of the (single) collision frequency $\beta$ instead of the collision frequencies $\beta_{\eta}^{\text{BE}}$ for $\eta$ and $\beta_{\kappa}^{\text{BE}}$ for the coefficients $\kappa$ and $\mu$. Thus, to optimize the agreement with the Boltzmann results for single component granular gases for the transport coefficients, one could consider $\beta$ as a free parameter of the model to reproduce either $\eta$ (if $\beta=\beta_{\eta}^{\text{BE}}$) or $\kappa$ and $\mu$ (if $\beta=\beta_{\kappa}^{\text{BE}}$).

Another possible source of discrepancy between the results derived from the Boltzmann equation and the kinetic model for a single granular gas could be the relevance of the non-Gaussian corrections (measured by the kurtosis $a_2$) to the Maxwellian distribution
\beq
\label{3.56.2}
f_\text{M}(\mathbf{V})=n\left(\frac{m}{2T}\right)^{3/2} \exp\left(-\frac{mV^2}{2T}\right).
\eeq
These corrections can play an important role in extreme inelasticity regimes. Although the Maxwellian distribution can be considered as a good approximation for the zeroth-order approximation $f^{(0)}$ in the region of thermal velocities (which is the relevant one for the lowest degree velocity moments of the first-order distribution $f^{(1)})$, significant differences between $f^{(0)}$ and $f_\text{M}$ are expected for strong inelasticity in the case of higher velocity moments (such as the pressure tensor and the heat flux). Since the single limiting case of the model \eqref{2.22} yields $a_2=0$, the discrepancies between the kinetic model and the Boltzmann equation for the transport coefficients $\eta$, $\kappa$, and $\mu$ are expected to increase with increasing dissipation, as Figs.\ \ref{fig8}a and \ref{fig9} clearly show.

Apart from the above two reasons, the deviations of the kinetic model results from the Boltzmann results for binary granular mixtures are also expected to increase with increasing the disparity in mass and/or diameter of the constituents of the system.}


\section{Uniform shear flow state}
\label{sec5}

The Chapman-Enskog solution of the inelastic Boltzmann equation for states with small spatial gradients is technically difficult but accessible. For more complex far from equilibrium states, the Boltzmann equation for granular mixtures becomes intractable. In these cases, kinetic models are used as a reliable alternative. Here, as a third problem in the paper, we study in this section the so-called simple or uniform shear flow (USF) state. Although this state has been extensively studied in the case of single component granular gases,\cite{LSJCh84,JR88,C89,HS92,LB94,GT96,SGN96,GS96,BRM97,ChR98,MGSB99,SGD04} the studies for granular mixtures are more scarce. \cite{Z95,AL02,CH02,AL03,MG02a,G03bis,L04} At a macroscopic level, the USF is defined by constant densities $n_i$, a uniform granular temperature $T$, and the linear velocity field \beq
\label{4.1}
U_{1,k}=U_{2,k}=U_k=a_{k\ell}r_\ell, \quad a_{k\ell}=a \delta_{kx}\delta_{\ell y},
\eeq
where $a$ is the \emph{constant} shear rate. At the microscopic level, one of the main advantages of the USF over other non-equilibrium problems is that in this state the spatial dependence of the distribution functions $f_i(\mathbf{r}, \mathbf{v}; t)$ arises only from their dependence on the peculiar velocity $V_k = v_k - a_{k \ell}r_\ell$. Thus, when the particle velocities are expressed in the Lagrangian frame moving with the flow velocity $U_k = a_{k \ell}r_\ell$, the USF becomes spatially homogeneous. This means that $f_i(\mathbf{r}, \mathbf{v}; t)=f_i(\mathbf{V}; t)$. This property is probably the main reason why this state has been studied extensively in molecular and granular gases, as it provides a clear framework for studying the nonlinear response of the system to strong shear.

However, the nature of the USF state is quite different for molecular and granular fluids, since in the latter a steady state is possible (without introducing external thermostats) when the viscous heating term is exactly compensated by the energy dissipated by collisions. Here we are mainly interested in obtaining the non-Newtonian transport properties of the mixture under steady conditions. In the steady state ($\partial_t f_i=0$), the set of kinetic equations \eqref{2.22} for the model reads
\beq
\label{4.2}
-2a V_y \frac{\partial f_1}{\partial V_x}+\omega_1 f_1-\epsilon_{1}
\frac{\partial}{\partial {\bf V}}\cdot (\mathbf{V}f_1)=\Phi_1,
\eeq
\beq
\label{4.2.0}
-2a V_y \frac{\partial f_2}{\partial V_x}+\omega_2 f_2-\epsilon_{2}
\frac{\partial}{\partial {\bf V}}\cdot (\mathbf{V}f_2)=\Phi_2.
\eeq

In the USF problem, the mass and heat fluxes vanish by symmetry reasons ($\mathbf{j}_1=\mathbf{q}=\mathbf{0}$) and the only nonzero flux is the pressure tensor $P_{k\ell}$. As a consequence, the relevant balance equation in the USF is that of the granular temperature $T$, Eq.\ \eqref{2.11}. In the steady state, Eq.\ \eqref{2.11} becomes
\beq
\label{4.3}
\frac{2}{3 n}a P_{xy}+\zeta T=0,
\eeq
where the pressure tensor $P_{k \ell}$ is defined by Eq.\ \eqref{2.13} and the expression of the cooling rate in the kinetic model is given by Eq.\ \eqref{2.8}. As mentioned earlier, there are two competing effects in the granular temperature equation according to Eq.\ \eqref{4.3}. On the one hand, the viscous heating term ($aP_{xy}<0$) causes the granular temperature to increase monotonically with time. On the other hand, since collisions are inelastic, there is a continuous loss of energy due to the collisional cooling term ($\zeta T>0$). In the steady state, these two effects cancel each other out. Due to the coupling between the shear stress $P_{xy}$ and the inelasticity (measured by the cooling rate $\zeta$), the reduced shear rate $a^* = a/\nu$ (where we recall that $\nu(T)\propto \sqrt{T}$) is only a function of the coefficients of restitution $\al_{ij}$ and the parameters of the mixture (mass and diameter ratios as well as concentration).

\subsection{Velocity moments in the steady USF}

As in the HCS, we are first interested in determining the velocity moments of the distributions $f_i(\mathbf{V})$. They are defined as
\beq
\label{4.4}
M_{k_1,k_2,k_3}^{(i)}=\int d\mathbf{V}\; V_x^{k_1}V_y^{k_2}V_z^{k_3}f_i(\mathbf{V}).
\eeq
The symmetry properties in the USF of the velocity distribution functions  $f_i(\mathbf{V})$ are \cite{GS03}
\beq
\label{4.5}
f_i(V_x,V_y,V_z)=f_i(V_x,V_y,-V_z),
\eeq
\beq
\label{4.5.1}
f_i(V_x,V_y,V_z)=f_i(-V_x,-V_y,V_z).
\eeq
According to these symmetry properties, the nonzero velocity moments are when $k_1+k_2$ and $k_3$ are even numbers.

Let us focus on the moments $M_{k_1,k_2,k_3}^{(1)}$ since the moments $M_{k_1,k_2,k_3}^{(2)}$ corresponding to the distribution of the species $2$ can be easily obtained from the former by making the change $1\leftrightarrow 2$. To get $M_{k_1,k_2,k_3}^{(1)}$  one multiplies both sides of Eq.\ \eqref{4.4} by $V_x^{k_1}V_y^{k_2}V_z^{k_3}$ and integrates over velocity to achieve the result
\beq
\label{4.6}
2 a k_1 M_{k_1-1,k_2+1,k_3}^{(1)}+(\omega_1+k\epsilon_1)M_{k_1,k_2,k_3}^{(1)}=N_{k_1,k_2,k_3}^{(1)},
\eeq
where
\beqa
\label{4.7}
N_{k_1,k_2,k_3}^{(1)}&=&\int d\mathbf{V}\; V_x^{k_1}V_y^{k_2}V_z^{k_3}\Phi_1(\mathbf{V})=n_1 \Gamma_{k_1,k_2,k_3}
\nonumber\\
& &\times
\Bigg[\omega_{11}\left(\frac{2T_1}{m_1}\right)^{k/2}+
\omega_{12}\left(\frac{2T_{12}}{m_1}\right)^{k/2}\Bigg].\nonumber\\
\eeqa
Here, we recall that $\Gamma_{k_1,k_2,k_3}$ is defined in Eq.\ \eqref{3.11}. The solution to Eq.\ \eqref{4.6} can be written as (see the Appendix A of Ref.\ \onlinecite{GL95})
\beq
\label{4.14}
M_{k_1,k_2,k_3}^{(1)}=\sum_{q=0}^{k_1}\frac{k_1!}{(k_1-q)!}\frac{(-2a)^q}{(\omega_1+k\epsilon_1)^{1+q}}N_{k_1-q,k_2+q,k_3}.
\eeq

Equation \eqref{4.14} is still a formal expression as we do not know the dependence of both the temperature ratio $\gamma=T_1/T_2$ and the (reduced) shear rate $a^*$ on the coefficients of restitution and the parameters of the mixture. To determine these quantities, one can consider, for example, the dimensionless version of Eq.\ \eqref{4.3}, which leads to the relation
\beq
\label{4.16}
a^*=-\frac{3}{2}\frac{x_1\gamma_1 \zeta_1^*+x_2\gamma_2 \zeta_2^*}{x_1 P_{1,xy}^*+x_2 P_{2,xy}^*},
\eeq
where $\zeta_i^*=\zeta_i/\nu$, $P_{i,k \ell}^*=P_{i,k \ell}/(x_i p)$, and
\beq
\label{4.17}
\gamma_1=\frac{T_1}{T}=\frac{\gamma}{x_2+x_1\gamma}, \quad
\gamma_2=\frac{T_2}{T}=\frac{1}{x_2+x_1\gamma}.
\eeq
Finally, for $i=1$ and 2, the requirements
\beq
\label{4.17.1}
M_{2,0,0}^{(i)}+M_{0,2,0}^{(i)}+M_{0,0,2}^{(i)}=3\frac{n_iT_i}{m_i},
\eeq
yield the condition
\beq
\label{4.18}
\frac{M_{2,0,0}^{(1)}+M_{0,2,0}^{(1)}+M_{0,0,2}^{(1)}}{M_{2,0,0}^{(2)}+M_{0,2,0}^{(2)}+M_{0,0,2}^{(2)}}=\frac{x_1}{x_2}\theta^{-1},
\eeq
where we recall that $\theta=m_1T_2/m_2T_1$. For given values of the parameter space of the mixture, the numerical solution to Eq\ \eqref{4.18} gives the temperature ratio $\gamma$. Once $\gamma$ is known, Eq.\ \eqref{4.16} gives $a^*$, while the explicit dependence of the (dimensionless) moments on the parameters of the mixture is given by Eq.\ \eqref{4.14}.

\begin{figure}[h]
\includegraphics[width=0.4\textwidth]{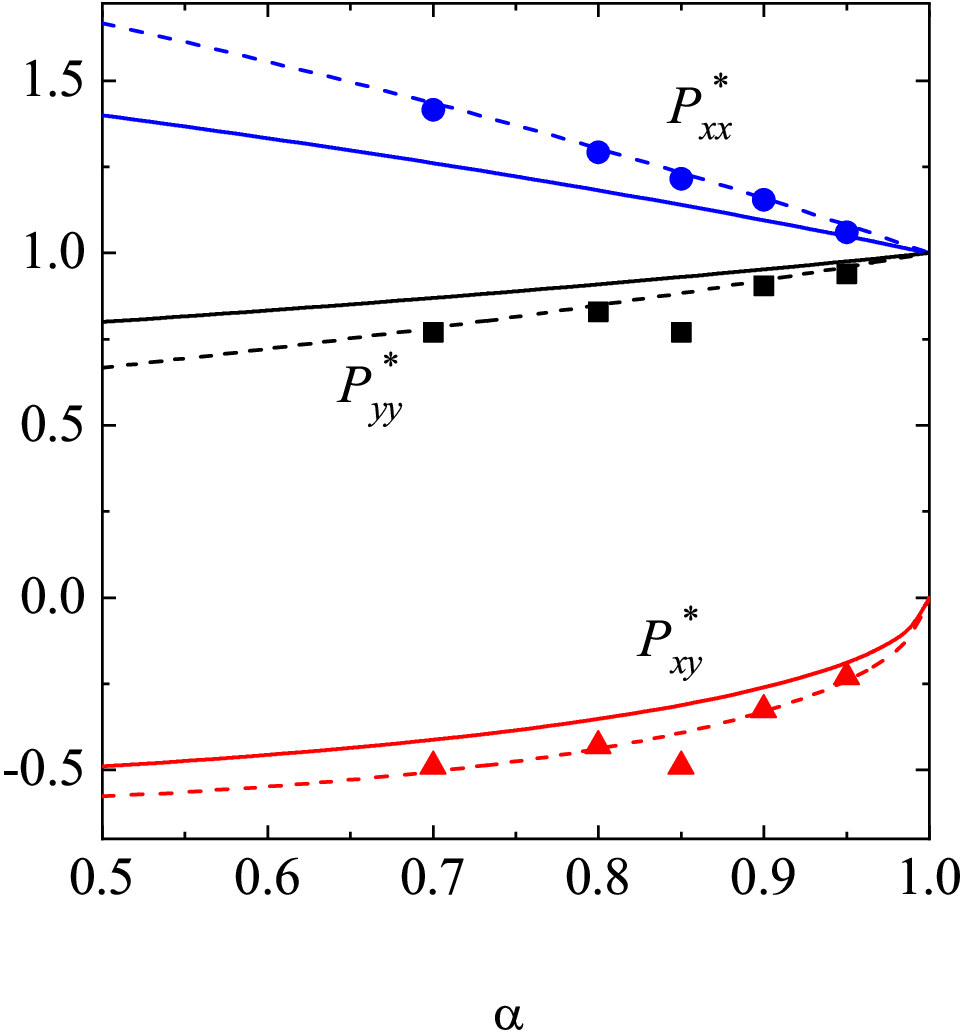}
\caption{Plot of the (reduced) elements of the pressure tensor as functions of the coefficient of restitution $\al$ for a monocomponent granular gas. The solid lines are the results derived here from the kinetic model while the dashed lines correspond to the results obtained by solving approximately the Boltzmann equation by means of Grad's moment method. \cite{G02,SGD04} Symbols refer to Monte Carlo simulation results obtained in Ref.\ \onlinecite{MG02a}.
}
\label{fig10}
\end{figure}
\begin{figure}[h!]
\includegraphics[width=0.4\textwidth]{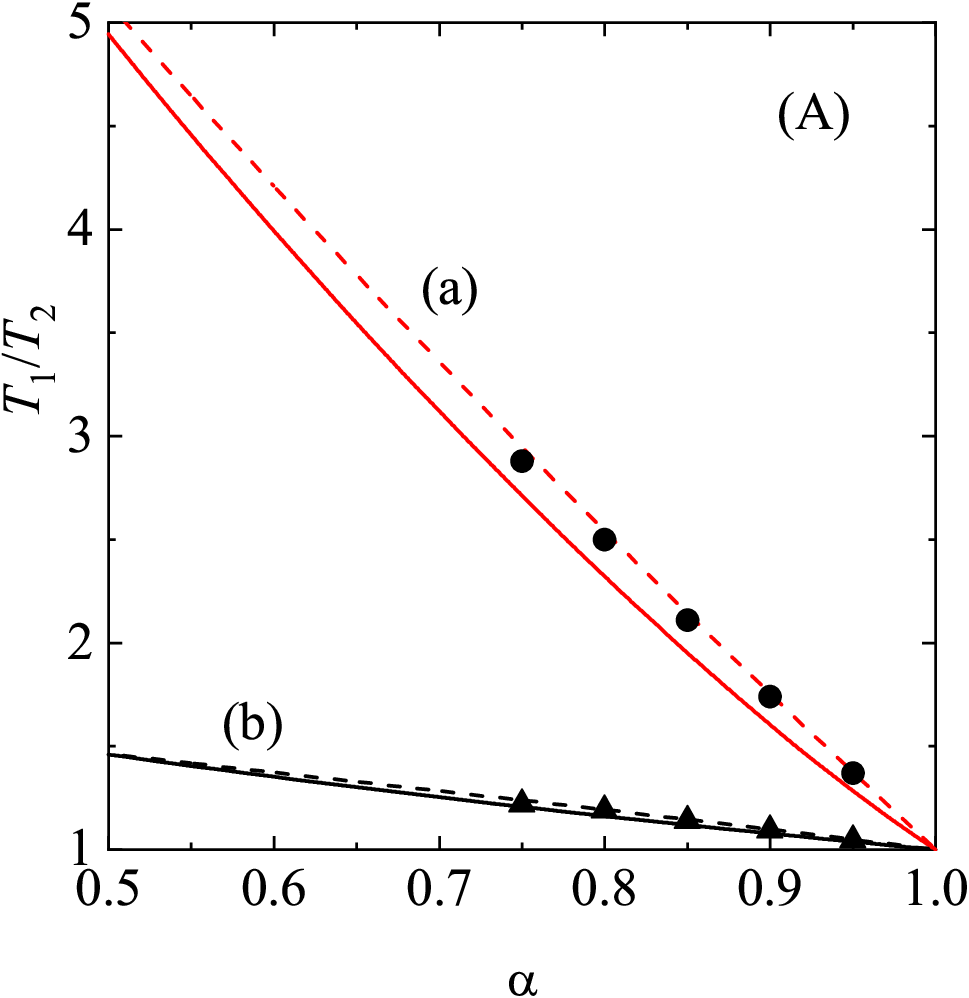}
\includegraphics[width=0.4 \textwidth,angle=0]{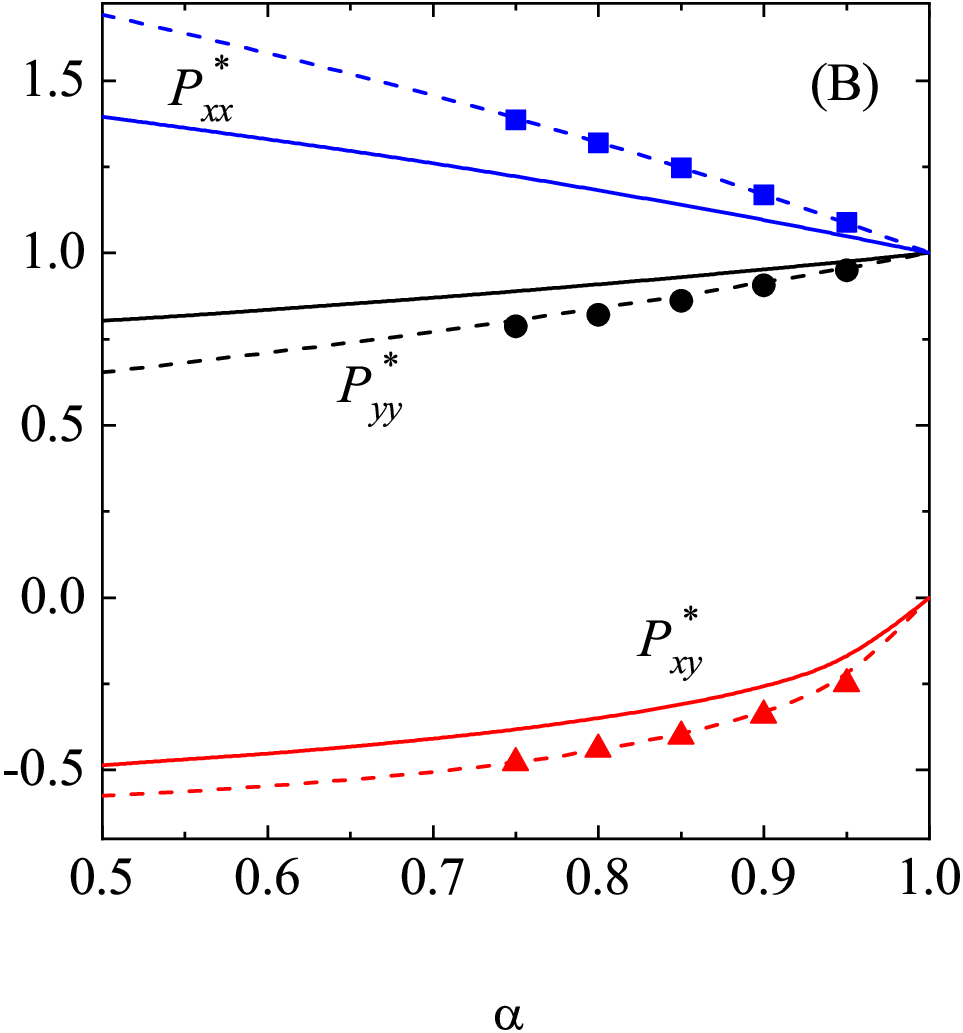}
\caption{Panel (A): Plot of the temperature ratio $T_1/T_2$ as a function of the (common) coefficient of restitution $\al$ for a binary granular mixture with $x_1=\frac{1}{2}$, $\sigma_1/\sigma_2=1$, and two different values of the mass ratio $m_1/m_2$: $m_1/m_2=10$ (a) and $m_1/m_2=2$ (b). The solid and dashed lines correspond to the results obtained here from the kinetic model and the Boltzmann equation, respectively. Symbols are DSMC simulations reported in Ref.\ \onlinecite{MG02a}. Panel (B): Plot of the (reduced) elements of the pressure tensor as functions of the (common) coefficient of restitution $\al$ for a binary  granular mixture with $x_1=\frac{1}{2}$, $\sigma_1/\sigma_2=1$, and $m_1/m_2=2$. The solid and dashed lines correspond to the results obtained here from the kinetic model and the Boltzmann equation, respectively. Symbols refer to Monte Carlo simulation results obtained in Ref.\ \onlinecite{MG02a}.}
\label{fig11}
\end{figure}

\subsection{Rheological properties}

The most relevant moments are those related with the nonzero elements of the pressure tensor. From Eq.\ \eqref{4.14}, one gets the results
\beq
\label{4.19}
P_{1,yy}^*=P_{1,zz}^*=\frac{\omega_{11}^*\gamma_1
+\omega_{12}^*\gamma_{12}}{\omega_1^*+2\epsilon_1^*}, \quad
P_{1,xx}^*=3\gamma_1-2P_{1,yy}^*,
\eeq
\beq
\label{4.19.1}
P_{1,xy}^*=-\frac{2P_{1,yy}^*}{\omega_1^*+2\epsilon_1^*}a^*.
\eeq
Here, $\gamma_{12}=T_{12}/T=\gamma_1+2\mu_{12}\mu_{21}(\gamma_2-\gamma_1)$. The expression of $P_{2,k\ell}^*$ can be easily derived from Eq.\ \eqref{4.19} by the change $1\leftrightarrow 2$. The (reduced) pressure tensor of the mixture $P_{k\ell}^*=P_{k\ell}/p$ is
\beq
\label{4.20}
P_{k\ell}^*=x_1 P_{1,k\ell}^*+x_2 P_{2,k\ell}^*.
\eeq

For mechanically equivalent particles, $\gamma_1=\gamma_2=\gamma_{12}=1$ and Eqs.\ \eqref{4.19} and \eqref{4.19.1} yield
\beq
\label{4.21}
P_{yy}^*=\frac{2}{3-\al}, \quad P_{xy}^*=-\frac{3}{\sqrt{2\pi}} \frac{a^*}{(1+\al)(3-\al)^2}.
\eeq
The $\al$-dependence of the (reduced) shear rate $a^*$ for a monocomponent dilute granular gas is obtained from Eqs.\ \eqref{4.16} and \eqref{4.21}:
\beq
\label{4.22}
a^{*2}=\frac{2}{3}\pi(3-\al)^2(1+\al)^2(1-\al).
\eeq
Figure \ref{fig10} shows the $\al$-dependence of the (reduced) elements of the pressure tensor, $P_{xx}^*$, $P_{yy}^*$, and $P_{xy}^*$, for a single component granular gas under USF. The results obtained here in Eq.\ \eqref{4.21} are compared with the approximate results of Refs.\ \onlinecite{G02, SGD04}, which were derived by solving the Boltzmann equation using Grad's moment method. \cite{G49} For completeness, numerical solutions \cite{BRM97,MG02a} of the Boltzmann equation using the DSMC method \cite{B94} are also shown. Both the Boltzmann equation and the kinetic model clearly predict anisotropy in the diagonal elements of the pressure tensor in the shear plane ($P_{xx}^* \neq P_{yy}^*$, but $P_{yy}^* = P_{zz}^*$). It is important to indicate that the simulations also show anisotropy in the plane orthogonal to the flow velocity; in fact, $P_{zz}^*$ is slightly larger than $P_{yy}^*$. However, the difference $P_{zz}^*-P_{yy}^*$ is generally small and tends to zero as the inelasticity decreases in collisions. \cite{BRM97,MG02a}

We observe in Fig.\ \ref{fig10} that the predictions of the kinetic model are in qualitative agreement with the results of the DSMC, although there are some quantitative differences, especially under strong dissipation for $P_{xx}^*$. The approximate results derived from the Boltzmann equation show better agreement with computer simulations than those from the kinetic model. As mentioned in subsection \ref{sec3.B},  the kinetic model results could be improved (as shown, for example, in Fig.\ 5 of Ref.\ \onlinecite{BRM97}) by adjusting the effective collision frequency in Brey \emph{et al.}'s original kinetic model \cite{BDS99} (which can be treated as a free parameter) to match the Boltzmann value for the Navier-Stokes shear viscosity for elastic collisions. However, since the expressions in Eq.\ \eqref{4.21} are obtained by solving the set of equations \eqref{4.2}--\eqref{4.2.0} in the limiting case of identical particles, the model has no free parameters, since the collision frequencies $\nu_{ij}$ are defined by Eq.\ \eqref{2.17}.

Complementing the results shown in Fig.\ \ref{fig10}, panels (A) and (B) of Fig.\ \ref{fig11} show the dependence of both the temperature ratio $T_1/T_2$ and the (reduced) elements of the pressure tensor, respectively, on the (common) coefficient of restitution $\al$ for different mixtures. The theoretical results derived here from the kinetic model are compared with those obtained by solving the Boltzmann equation by means of Grad's moment method \cite{G02,SGD04} and Monte Carlo simulations. \cite{MG02a} Similar to Fig.\ \ref{fig10}, there is reasonably good agreement between the Boltzmann and kinetic model results, especially for the temperature ratio. Also, as in the case of single component gases, the approximate Boltzmann results show better agreement with simulations than those from the kinetic model. A comparison between Fig.\ \ref{fig10} and panel (B) of Fig.\ \ref{fig11} clearly shows the weak influence of the mass ratio on the rheological properties of the system.

\subsection{Velocity distribution function in the USF}

\begin{figure}[h]
\includegraphics[width=0.4\textwidth]{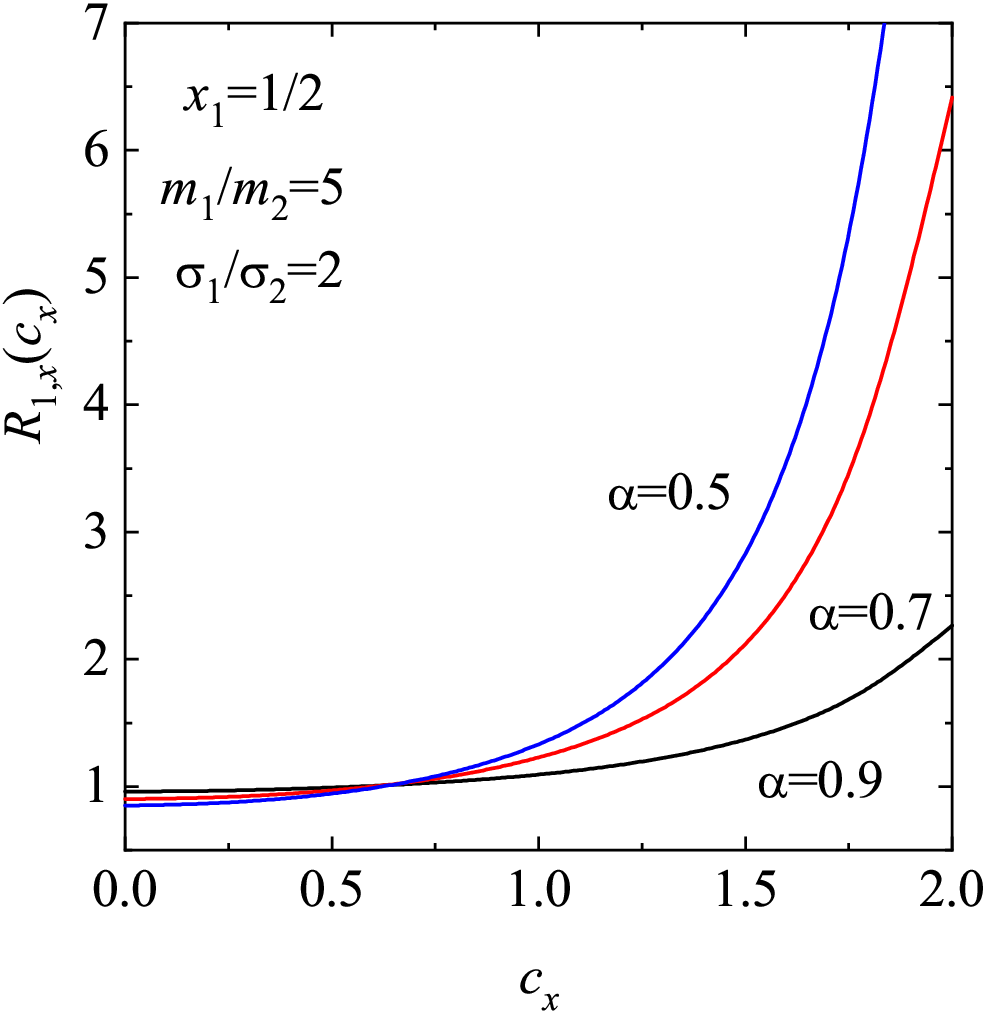}
\caption{Plot of the ratio $R_{1,x}(c_x)=\varphi_{1,x}(c_x)/\varphi_{1,x}^\text{el}(c_x)$ versus the (scaled) velocity $c_x$ for $x_1=\frac{1}{2}$, $m_1/m_2=5$, $\sigma_1/\sigma_2=2$, and three different values of the (common) coefficient of restitution $\al_{ij}=\al$: $\al=0.9$, 0.7 and 0.5. The scaled USF distribution $\varphi_{1,x}$ is given by Eq.\ \eqref{4.29}.}
\label{fig12}
\end{figure}

As in the case of the HCS, the explicit forms of the velocity distribution functions $f_i(\mathbf{V})$ of the granular binary mixture under USF can be also obtained. Let us focus on the velocity distribution $f_1(\mathbf{V})$ of species 1. Equation \eqref{4.2} can be cast into the form
\beq
\label{4.23}
\Big(\omega_1-3\epsilon_1-\epsilon_1 \mathbf{V}\cdot \frac{\partial}{\partial \mathbf{V}}-2a V_y \frac{\partial}{\partial V_x}\Big)f_1(\mathbf{V})=\Phi_1(\mathbf{V}).
\eeq
A formal (hydrodynamic) solution to Eq.\ \eqref{4.23} is
\beqa
\label{4.24}
f_1(\mathbf{V})&=&\Big(\omega_1-3\epsilon_1-\epsilon_1 \mathbf{V}\cdot \frac{\partial}{\partial \mathbf{V}}-2a V_y \frac{\partial}{\partial V_x}\Big)^{-1}\Phi_1(\mathbf{V})\nonumber\\
&=&\int_0^\infty\; ds\; e^{-(\omega_1-3\epsilon_1)s}e^{\epsilon_1 s\mathbf{V}\cdot \frac{\partial}{\partial \mathbf{V}}} e^{2 a s V_y \frac{\partial}{\partial V_x}}\Phi_1(\mathbf{v}).\nonumber\\
\eeqa
The action of the shift operators $e^{\epsilon_1 s\mathbf{v}\cdot \frac{\partial}{\partial \mathbf{v}}}$ and $e^{2 a s V_y \frac{\partial}{\partial V_x}}$ in velocity space on an arbitrary function $g(\mathbf{V})$ is
\beq
\label{4.25}
e^{\epsilon_1 s\mathbf{V}\cdot \frac{\partial}{\partial \mathbf{V}}}g(V_x,V_y,V_z)=g(e^{\epsilon_1 s}V_x, e^{\epsilon_1 s}V_y, e^{\epsilon_1 s}V_z),
\eeq
\beq
\label{4.26}
e^{2 a s V_y \frac{\partial}{\partial V_x}}g(V_x,V_y,V_z)=g(V_x+2as V_y,V_y,V_z).
\eeq
\vicente{
\begin{widetext}
Taking into account Eqs.\ \eqref{4.25} and \eqref{4.26}, the velocity distribution function $f_1(\mathbf{V})$ can be written as
\beqa
\label{4.26.1}
f_1(\mathbf{V})&=&\int_0^\infty\; ds\; e^{-(\omega_1-3\epsilon_1)s}n_1
\Bigg\{\omega_{11}\Big(\frac{m_1}{2T_1}\Big)^{3/2}\exp\Bigg[-\frac{m_1}{2T_1}e^{2\epsilon_1 s}\Big(V^2+4asV_xV_y+4a^2s^2V_y^2\Big)\Bigg]\nonumber\\
& &
+\omega_{12}\Big(\frac{m_1}{2T_{12}}\Big)^{3/2}\exp\Bigg[-\frac{m_1}{2T_{12}}e^{2\epsilon_1 s}\Big(V^2+4a sV_xV_y+4a^2s^2V_y^2\Big)\Bigg]\Bigg\}.
\eeqa
\end{widetext}

As in the study of the HCS, it is convenient to express $f_1(\mathbf{V})$ in dimensionless form by introducing the dimensionless quantities $\tau=\nu s$, $\omega_1^*$, $\epsilon_1^*$, $\theta_1$, $\theta_{12}$, and $\mathbf{c}$. Thus, the velocity distribution function of species 1 in the USF problem can be cast into the form}
\beq
\label{4.27}
f_1(\mathbf{V})= n_1 v_\text{th}^{-3} \varphi_1 (\mathbf{c}),
\eeq
where the scaled distribution $\varphi_1 (\mathbf{c})$ is
\begin{widetext}
\beqa
\label{4.28}
\varphi_1 (\mathbf{c})&=&\pi^{-3/2}\int_0^\infty\; d\tau\; e^{-(\omega_1^*-3\epsilon_1^*)\tau}\Bigg\{\omega_{11}^*\theta_1^{3/2}
\exp\Big[-\theta_1 e^{2\epsilon_1^* \tau}\left(c^2+4a^* \tau c_x c_y+4 a^{*2}\tau^2 c_y^2\right)\Big]\nonumber\\
& & +\omega_{12}^*\theta_{12}^{3/2}
\exp\Big[-\theta_{12} e^{2\epsilon_1^* \tau}\left(c^2+4a^* \tau c_x c_y+4 a^{*2}\tau^2 c_y^2\right)\Big]\Bigg\}.
\eeqa
\end{widetext}
It can be checked (see the Appendix \ref{appA} for some technical details) that the expression \eqref{4.28} reproduces the moments \eqref{4.14}. This agreements confirms the consistency of the results reported in this section for the USF problem.

To illustrate the dependence of $\varphi_1(\mathbf{c})$ on the (dimensionless) velocity $\mathbf{c}$, we define the marginal distribution
\beq
\label{4.29}
\varphi_{1,x}(c_x)=\int_{-\infty}^{+\infty}dc_y\int_{-\infty}^{+\infty}dc_z\; \varphi_1(\mathbf{c}).
\eeq
Substituting Eq.\ \eqref{4.28} into Eq.\ \eqref{4.29} and performing the velocity integrals one gets
\begin{widetext}
\beq
\label{4.29.1}
\varphi_{1,x}(c_x)=
\pi^{-1/2}\int_0^\infty d\tau\;
\frac{e^{-(\omega_1^*-\epsilon_1^*)\tau}}{\sqrt{1+4a^{*2}\tau^2}} \Bigg[\omega_{11}^*\theta_1^{1/2}
\exp\Bigg(-\theta_1 e^{2\epsilon_1^*\tau}\frac{c_x^2}{1+4a^{*2}\tau^2}\Bigg)+\omega_{12}^*\theta_{12}^{1/2}
\exp\Bigg(-\theta_{12} e^{2\epsilon_1^*\tau}\frac{c_x^2}{1+4a^{*2}\tau^2}\Bigg)\Bigg].
\eeq
\end{widetext}
For elastic collisions, $\epsilon_1^*=0$, $a^*=0$, $\theta_1=\theta_{12}=2\mu_{12}$ and $\varphi_{1,x}(c_x)$ reduces to the equilibrium distribution \eqref{3.27}, as expected.

Figure \ref{fig12} shows the ratio $R_{1,x}(c_x)=\varphi_{1,x}(c_x)/\varphi_{1,x}^\text{el}(c_x)$ as a function of the (scaled) velocity $c_x$ for the binary mixtures with $x_1=\frac{1}{2}$, $m_1/m_2=5$, $\sigma_1/\sigma_2=2$, and three different values of the (common) coefficient of restitution $\al_{ij}=\al$: $\al=0. 9$, 0.7 and 0.5. As in the case of HCS, we observe quite a distortion of the USF distribution $\varphi_{1,x}(c_x)$ with respect to its equilibrium value $\varphi_{1,x}^\text{el}(c_x)$. The deviation of $\varphi_{1,x}(c_x)$ from $\varphi_{1,x}^\text{el}(c_x)$ increases with increasing dissipation. Furthermore, a comparison with the (marginal) distribution of the HCS (see Fig.\ \ref{fig6}) shows that the growth of $R_{1,x}(c_x)$ with $c_x$ is more pronounced in the USF than in the HCS. Thus, although the collisional cooling effect (measured by the term $\zeta T$) is balanced by the viscous heating effect (measured by the term $-a P_{xy}$) in the (steady) USF state, for large velocities the particle population (relative to its elastic value) is larger in the USF state than in the HCS. Finally, it must also be remembered that in a kinetic model one cannot expect to be able to accurately describe the population of particles whose velocities are beyond the thermal one, since the evolution of the distributions $f_i(\mathbf{V})$ is essentially given only in terms of the first five velocity moments (in the USF in terms of the partial temperatures $T_1$ and $T_2$).

\section{Discussion}
\label{sec6}

The determination of transport properties of multicomponent mixtures from the Boltzmann equation is in general a rather complicated problem. Because of these technical difficulties, researchers have usually considered kinetic model equations in which the Boltzmann collision operators $J_{ij}[f_i,f_j]$ are replaced by terms that retain the relevant physical properties of these operators but are mathematically simpler. This procedure has been widely used in the past years in the case of molecular mixtures, where several models \cite{GK56,S62,H65,H66,GS67,GSB89,AAP02,HHKPW21} have been proposed to obtain explicit expressions for the transport coefficients of the mixture. However, much fewer models have been proposed for granular mixtures (mixtures of mechanically different hard spheres undergoing inelastic collisions). In fact, as mentioned in Sec.\ \ref{sec1}, we are only aware of the kinetic model proposed by Vega Reyes \textit{et al.} \cite{VGS07} for a mixture of inelastic hard spheres. This model is inspired in the equivalence between a gas of elastic hard spheres subject to a drag force with a gas of IHS. \cite{SA05} In this paper we have considered the model of Vega Reyes \textit{et al.} \cite{VGS07} where the elastic Boltzmann collision operators present in the original model are replaced by the relaxation terms of the well-known GK model for molecular mixtures. \cite{GK56} In this context, the kinetic model used here can be considered as a natural extension of the GK model to granular mixtures.

Three different non-equilibrium situations were considered. As a first step, the HCS was analyzed. The study of this state is crucial for the determination of the Navier-Stokes transport coefficients of the mixture, since the local version of the HCS plays the role of the reference state in the Chapman-Enskog perturbative method. \cite{CC70,G19} Surprisingly, depending on the parameters of the mixture, our study of the relaxation of the velocity moments to their HCS forms has shown the possible divergence of these moments, especially for sufficiently high degree moments. This kind of divergence could question the validity of a normal (or hydrodynamic) solution of the Boltzmann equation in the HCS. Once the HCS is well characterized, as a second problem we have obtained the exact forms of the Navier-Stokes transport coefficients in terms of the parameter space of the system. Finally, as a third problem, the rheological properties of a sheared granular mixture have also been derived.

The use of the kinetic model has allowed not only to obtain the exact forms of the linear and nonlinear transport properties of the mixture, but also to obtain the explicit forms of the velocity distribution functions. This is one of the main advantages of considering a kinetic model of the Boltzmann equation.

Comparison with both the (approximate) theoretical results of the Boltzmann equation and computer simulations shows in some cases an excellent agreement (temperature ratio in the HCS and the diffusion transport coefficients), in others a reasonable quantitative agreement (the Navier--Sokes shear viscosity and the rheological properties), while more significant discrepancies are present in the case of the heat flux transport coefficients. Regarding the velocity distribution functions, based on previous comparisons with DSMC results, \cite{BRM97} it is expected that the model gives accurate results for small velocities, but important differences are likely to appear in the high-velocity region. We hope that the present paper will stimulate the performance of these simulations to confirm the above expectations.

\vicente{As discussed in subsection \ref{sec3.A}, one of the surprising results derived from the kinetic model in the HCS is the divergence of high-degree velocity moments under certain parameter conditions. On the other hand, in contrast to the results recently obtained for IMM in the HCS, \cite{SG23} this kind of divergence appears for very high degree moments (see for example Fig.\ \ref{fig1}). Thus, this unphysical behavior precludes the failure of a hydrodynamic description of the granular mixture from the kinetic model. As mentioned before, the singular behavior of these high degree moments could have some implications on the existence of the (scaled) HCS solution \eqref{3.8} for values of the coefficient of restitution smaller than the critical value $\al_c$. Needless to say, elucidation of this point requires computer simulations of the Boltzmann equation \cite{B94} to clarify whether such a divergence is also present in the original Boltzmann equation or whether it is actually a drawback of the kinetic model.}

\vicente{Another approach for studying transport in granular mixtures different from the one followed in this paper is the possibility of establishing a linear relationship between driving forces and moment matrices through a large resistance matrix. This type of approach has recently been used to model macrotransport processes of elongated microswimmers involving anisotropic diffusion. \cite{GJTChL24} In the case of non-equilibrium steady states for granular mixtures, from a kinematic point of view, the introduction of non-zero, albeit small, driving forces could be considered as an interesting alternative to the use of kinetic models.}

In conclusion, the results reported here can be considered as a testimony of the reliability of the kinetic model \eqref{2.22} for the study of nonequilibrium problems where the use of the original Boltzmann equation turns out to be unapproachable. In particular, the derivation of the transport coefficients of a granular binary mixture characterizing the
transport around the USF is an interesting project for the near future. Given the technical difficulties involved in such a calculation, the kinetic model \eqref{2.22} can be considered as a useful starting point.


\acknowledgments

The work of V.G. is supported from Grant No. PID2020-112936GB-I00 funded by MCIN/AEI/ 10.13039/501100011033.
\vspace{0.5cm}

\textbf{AUTHOR DECLARATIONS}\\
\textbf{Conflict of Interest}\\
The authors have no conflicts to disclose.

\vspace{0.5cm}
\textbf{Author Contributions} \\
\textbf{Pablo Avil\'es}: Formal analysis (equal); Investigation (equal); Software (equal); Writing–review\&editing (equal).
\textbf{David Gonz\'alez M\'endez}: Formal analysis (equal); Investigation (equal); Software (equal); Writing–review\&editing (equal).
\textbf{Vicente Garz\'o}: Formal analysis (equal); Investigation (equal); Writing/Original Draft Preparation (lead); Writing–review\&editing(equal).
\vspace{0.5cm}

\textbf{DATA AVAILABILITY}\\

The data that support the findings of this study are available from the corresponding author upon reasonable request.

\appendix
\section{Velocity moments from the distribution functions}
\label{appA}

\begin{widetext}
In this Appendix we obtain the expressions of the velocity moments of the HCS and the USF problems from their corresponding velocity distribution functions .

Let us start with the HCS. The dimensionless moments $M_{k_1,k_2,k_3}^{*(1)}$ are defined as
\beq
\label{a1}
M_{k_1,k_2,k_3}^{*(1)}=\int d\mathbf{c}\; c_x^{k_1}c_y^{k_2}c_z^{k_3}\varphi_1(\mathbf{c}),
\eeq
where we recall that $\mathbf{c}=\mathbf{v}/v_\text{th}$ and $\varphi_1(\mathbf{c})$ in the HCS is given by Eq.\ \eqref{3.25}. Substitution of Eq.\eqref{3.25} into Eq.\ \eqref{a1} yields
\beqa
\label{a2}
M_{k_1,k_2,k_3}^{*(1)}&=&\pi^{-3/2}\int_0^\infty\; d\tau\; e^{-(\omega_1^*-3\xi_1^*)\tau}\int d\mathbf{c}\; c_x^{k_1}c_y^{k_2}c_z^{k_3}\Big[\omega_{11}^*\theta_1^{3/2}
\exp\Big(-\theta_1 e^{2\xi_1^*\tau}c^2\Big)+\omega_{12}^*\theta_{12}^{3/2}
\exp\Big(-\theta_{12} e^{2\xi_1^*\tau}c^2\Big)\Big]\nonumber\\
&=&\pi^{-3/2}\int_0^\infty\; d\tau\; e^{-(\omega_1^*-3\xi_1^*)\tau}\Big[\omega_{11}^*\theta_1^{-k/2}e^{-(k+3)\xi_1^*\tau}
+\omega_{12}^*\theta_{12}^{-k/2}e^{-(k+3)\xi_1^*\tau}\Big]\left(\int d\mathbf{\varpi}\; \varpi_x^{k_1}\varpi_y^{k_2}\varpi_z^{k_3}e^{-\varpi^2}\right)\nonumber\\
&=&\Gamma_{k_1,k_2,k_3}
\left(\omega_{11}^*\theta_1^{-k/2}+\omega_{12}^*\theta_{12}^{-k/2}\right)\int_0^\infty\; d\tau\; e^{-(\omega_1^*+k\xi_1^*)\tau},
\eeqa
where $\Gamma_{k_1,k_2,k_3}$ is defined by Eq.\ \eqref{3.11}.
The integral over $\tau$ in the third line of Eq.\ \eqref{a2} is finite if
\beq
\label{a3}
\omega_1^*+k\xi_1^*>0.
\eeq
As shown in subsection \ref{sec3.A}, the inequality \eqref{a3} gives the condition for which the dimensionless moments in the HCS are convergent. According to the expression \eqref{3.21} of $\xi_1^*$, the condition \eqref{a3} can be written more explicitly as
 \beq
\label{a3.0}
k<\frac{x_1\left(\frac{\sigma_1}{\sigma_{12}}\right)^2\sqrt{\frac{2}{\theta_1}}(1+\al_{11})+x_2\sqrt{\frac{\theta_1+\theta_2}{\theta_1\theta_2}}(1+\al_{12})}{x_2\mu_{12}\mu_{21}\sqrt{\frac{\theta_1+\theta_2}{\theta_1\theta_2}}\left(\frac{T_1-T_2}{T_1}\right)(1+\al_{12})}.
\eeq
If the condition \eqref{a3.0} holds, then
Eq.\ \eqref{a2} leads to the result
\beq
\label{a4}
M_{k_1,k_2,k_3}^{*(1)}=\Gamma_{k_1,k_2,k_3}\frac{\omega_{11}^*\theta_1^{-k/2}+
\omega_{12}^*\theta_{12}^{-k/2}}{\omega_1^*+k\xi_1^*}.
\eeq
Equation \eqref{a4} agrees with Eq.\ \eqref{3.16}.

In the case of the USF, to get the (dimensionless) velocity moments $M_{k_1,k_2,k_3}^{*(1)}$ one has to take into account the property
\beq
\label{a5}
\int d\mathbf{V} g(V_x,V_y,V_z) e^{2 a s V_y \frac{\partial}{\partial V_x}} h(V_x,V_y,V_z)=\int d\mathbf{V} g(V_x-2as V_y,V_y,V_z)h(V_x,V_y,V_z).
\eeq
Substitution of the form \eqref{4.28} of the USF distribution into the definition \eqref{a1} and taking into account Eq.\ \eqref{a5},  one achieves the result
\beqa
\label{a6}
M_{k_1,k_2,k_3}^{*(1)}&=&\pi^{-3/2}\int_0^\infty\; d\tau\; e^{-(\omega_1^*-3\epsilon_1^*)\tau}\int d\mathbf{c}(c_x-2a^*\tau c_y)^{k_1}c_y^{k_2}c_z^{k_3}\Bigg[\omega_{11}^*\theta_1^{3/2}
\exp\Big(-\theta_1 e^{2\epsilon_1^* \tau}c^2\Big)\nonumber\\
& & +\omega_{12}^*\theta_{12}^{3/2}
\exp\Big(-\theta_{12} e^{2\epsilon_1^* \tau}c^2\Big)\Bigg]\nonumber\\
&=&\sum_{q=0}^{k_1}\frac{k_1!}{q!(k_1-q)!}\Gamma_{k_1-q,k_2+q,k_3}
\int_0^\infty\; d\tau\; (-2a^*\tau)^q e^{-(\omega_1^*+k\epsilon_1^*)\tau}\left(\omega_{11}^*\theta_1^{-k/2}+\omega_{12}^*\theta_{12}^{-k/2}\right),
\eeqa
where $\Gamma_{k_1,k_2,k_3}$ is defined in Eq.\ \eqref{3.11} and in the last step we have expanded $(c_x-2a^*\tau c_y)^{k_1}$ and integrated over $\mathbf{c}$.  After performing the $\tau$-integration in Eq.\ \eqref{a6}, one finally gets the result
\beq
\label{a7}
M_{k_1,k_2,k_3}^{*(1)}=\sum_{q=0}^{k_1}\frac{k_1!}{(k_1-q)!}\Gamma_{k_1-q,k_2+q,k_3}\frac{(-2a^*)^q}{(\omega_1^*+k\epsilon_1^*)^{1+q}}
\left(\omega_{11}^*\theta_1^{-k/2}+\omega_{12}^*\theta_{12}^{-k/2}\right).
\eeq
Equation \eqref{a7} is identical to Eq.\ \eqref{4.14} when you write it in dimensionless form. This shows the consistency of our results.


\section{First-order distribution function. Mass, momentum, and heat fluxes}
\label{appB}

 In the first-order of the spatial gradients, the first-order distribution function $f_1^{(1)}$ verifies the kinetic equation
\beq
\label{b1}
\partial_t^{(0)}f_1^{(1)}+\frac{1}{2}\omega_1 f_1^{(1)}-\frac{1}{2}\epsilon_1\left(\frac{\partial}{\partial \mathbf{V}}\cdot \mathbf{V}f_1^{(1)}\right)
=\frac{1}{2}\left(\omega_{11}f_{11}^{(1)}+\omega_{12}f_{12}^{(1)}\right)-\left(D_t^{(1)}+\mathbf{V}\cdot \nabla\right)f_1^{(0)}-\frac{\epsilon_1}{2\rho_1}
\mathbf{j}_1^{(1)}\cdot \frac{\partial}{\partial \mathbf{V}} f_1^{(0)},
\eeq
where
\beq
\label{b2}
\mathbf{j}_1^{(1)}=\int d\mathbf{v} m_1 \mathbf{V} f_1^{(1)}
\eeq
is the first-order contribution to the mass flux,
\beq
\label{b3}
f_{11}^{(1)}(\mathbf{V})=\frac{f_{11}^{(0)}}{n_1 T_1} \mathbf{V}\cdot \mathbf{j}_1^{(1)}, \quad f_{12}^{(1)}(\mathbf{V})=\frac{\mu_{12}}{T_{12}}\frac{n_2-n_1}{n_1n_2}\mathbf{V}\cdot \mathbf{j}_1^{(1)} f_{12}^{(0)}(\mathbf{V}), \quad f_{ij}^{(0)}=n_i\left(\frac{m_i}{2\pi
T_{ij}}\right)^{d/2} \exp\left(-\frac{m_i}{2T_{ij}}V^2\right),
\eeq
and $D_t^{(1)}=\partial_t^{(1)}+\mathbf{U}\cdot \nabla$.

Given that the action of the operator $\left(D_t^{(1)}+\mathbf{V}\cdot \nabla\right)$ on the zeroth-order distribution $f_1^{(0)}$ is formally the same as in the original inelastic Boltzmann equation, \cite{GD02,GM07} we can omit part of the steps followed by the derivation of the kinetic equation of $f_1^{(1)}$. We refer to the interested reader to the Appendix A of Ref.\ \onlinecite{GM07} for more specific details. The kinetic equation for the first-order distribution function $f_1^{(1)}(\mathbf{V})$ is
\beqa
\label{b5}
& & \partial_t^{(0)}f_1^{(1)}+\frac{1}{2}\omega_1 f_1^{(1)}-\frac{1}{2}\epsilon_{1}
\frac{\partial}{\partial \mathbf{V}}\cdot \left(\mathbf{V} f_1^{(1)}\right)=\mathbf{A}_1\cdot \nabla x_1+\mathbf{B}_1\cdot \nabla p+\mathbf{C}_1\cdot \nabla T+D_{1,ij}\nabla_i U_j\nonumber\\
& &+\mathbf{j}_1^{(1)}\cdot \left[\left({\omega}_{11} \frac{f_{11}^{(0)}}{2n_1 T_1}+{\omega}_{12}\frac{\mu_{12}}{2T_{12}}\frac{n_2-n_1}{n_1n_2}f_{12}^{(0)}\right)\mathbf{V}-
\frac{\epsilon_{1}}{2\rho_1}\frac{\partial}{\partial \mathbf{V}} f_1^{(0)}\right],
\eeqa
where
\beq
\label{b6}
\mathbf{A}_1(\mathbf{V})=\left(-\frac{\partial f_1^{(0)}}{\partial x_1}\right)_{p,T} \mathbf{V},\quad \mathbf{B}_1(\mathbf{V})=-\frac{1}{p}\left[f_1^{(0)}\mathbf{V}+\frac{p}{\rho}\left(\frac{\partial f_1^{(0)}}{\partial \mathbf{V}}\right)\right],
\eeq
\beq
\label{b8}
\mathbf{C}_1(\mathbf{V})=\frac{1}{T}\left[f_1^{(0)}+\frac{1}{2}\frac{\partial}{\partial \mathbf{V}}\cdot \left(\mathbf{V}f_1^{(0)}\right)\right]\mathbf{V},
\quad D_{1,ij}(\mathbf{V})=V_i\frac{\partial f_1^{(0)}}{\partial V_j}-\frac{1}{d}\delta_{ij}\mathbf{V}\cdot \frac{\partial f_1^{(0)}}{\partial \mathbf{V}}.
\eeq

The mass, momentum, and heat fluxes can be directly determined from the kinetic equation \eqref{b5}. Let us consider the mass flux. To achieve it, one multiplies both sides of Eq.\ \eqref{b5} by $m_1\mathbf{V}$ and integrates over $\mathbf{V}$. After some algebra, one gets
\beq
\label{b10}
\left(\partial_t^{(0)}+\nu_D\right)\mathbf{j}_1^{(1)}=-p \left(\frac{\partial }{\partial x_1}x_1 \gamma_1\right)_{p,T}\nabla x_1-\left(x_1\gamma_1-\frac{\rho_1}{\rho}\right)\nabla p,
\eeq
where $\nu_D$ is defined by Eq.\ \eqref{3.40.1} and upon deriving Eq.\ \eqref{b10} use has been made of the constitutive equation \eqref{3.28} of the mass flux. Dimensional analysis shows that $D\propto T^{1/2}$, $D_p\propto T^{1/2}/p$, and $D_T\propto T^{1/2}$. Thus, taking into account the constitutive equation \eqref{3.28}, one has the result
\beqa
\label{b11}
\partial_t^{(0)}\mathbf{j}_1^{(1)}&=&-\zeta^{(0)}(T\partial_T+p\partial_p)\mathbf{j}_1^{(1)}
\nonumber\\
&=&\Big[\frac{m_1m_2 n}{2\rho}\zeta^{(0)}D+\rho(D_p+D_T)
\left(\frac{\partial \zeta ^{(0)}}{\partial x_{1}}\right)_{p,T}\Big]\nabla x_1
+
\frac{\rho \zeta^{(0)}}{p}\left(\frac{3}{2}D_p+D_T\right)\nabla p-\frac{\rho \zeta^{(0)}}{2T}D_p\nabla T,
\eeqa
where use has been made of the identities $\partial_t^{(0)}\nabla T=-\nabla (T\zeta^{(0)})$ and
$\partial_t^{(0)}\nabla p=-\nabla (p\zeta^{(0)})$. Inserting Eq.\ \eqref{b11} into Eq.\ \eqref{b10} allows to determine $D$, $D_p$, and $D_T$. Their expressions are given by Eqs.\ \eqref{3.38}--\eqref{3.40}.

The pressure tensor is given by Eq.\ \eqref{2.13} where the first-order contributions $P_{i,k\ell}^{(1)}$ are defined as
\beq
\label{b12}
P_{i,k\ell}^{(1)}=\int d\mathbf{v} m_i V_k V_\ell f_i^{(1)}(\mathbf{V}).
\eeq
We multiply both sides of Eq.\ \eqref{b5} (for $i=1,2$) by $m_i V_kV_\ell$ and integrates over velocity to get
\beq
\label{b13}
\left(\partial_t^{(0)}+\beta_i\right)P_{i,k\ell}^{(1)}=-n_i T_i \left(\nabla_\ell U_k+
\nabla_k U_\ell-\frac{2}{d}\delta_{k\ell}\nabla \cdot {\bf U}\right),
\eeq
where we recall that $\beta_i=(\omega_i+\epsilon_i)/2$.
The solution to Eq.\ \eqref{b13} has the form
\beq
\label{b14}
P_{i,k\ell}^{(1)}=-\eta_i \left(\nabla_\ell U_k+
\nabla_k U_\ell-\frac{2}{d}\delta_{k\ell}\nabla \cdot {\bf U}\right).
\eeq
Dimensionless analysis requires that $\eta_i\propto T^{1/2}$ so that, $\partial_t^{(0)} P_{i,k\ell}^{(1)}=-(\zeta^{(0)}/2)P_{i,k\ell}^{(1)}$. Substitution of this term into Eq.\ \eqref{b13} yields Eq.\ \eqref{3.42} for the shear viscosity coefficient $\eta=\eta_1+\eta_2$.

The first-order contribution to the heat flux $\mathbf{q}^{(1)}=\mathbf{q}_1^{(1)}+\mathbf{q}_2^{(1)}$, where
\beq
\label{b15}
\mathbf{q}_i^{(1)}=\int d\mathbf{v} \frac{m_i}{2} V^2 \mathbf{V}f_i^{(1)}(\mathbf{V}).
\eeq
As in the previous calculations, to achieve $\mathbf{q}_i^{(1)}$ we multiply both sides of Eq.\ \eqref{b5} (for $i=1,2$) by $\frac{1}{2}m_i V^2 \mathbf{V}$ and integrates over $\mathbf{V}$. The result is
\beq
\label{b16}
\left(\partial_t^{(0)}+\beta_i\right)\mathbf{q}_i^{(1)}=-\frac{5}{2}\frac{pT}{m_i}\Big(\frac{\partial}{\partial x_1}\left(x_i\gamma_i^2\right)\Big)_{p,T}\nabla x_1-\frac{5}{2}\frac{n_i T_i^2}{m_i p}\left(1-\frac{m_i p}{\rho T_i}\right)\nabla p-\frac{5}{2}\frac{n_i T_i^2}{m_i T}\nabla T+\frac{5}{4}A_i \mathbf{j}_i^{(1)},
\eeq
where we recall that $A_1$ is defined by Eq.\ \eqref{3.49} while $A_2$ can be easily obtained by interchanging $1\leftrightarrow 2$. According to the right-hand side of Eq.\ \eqref{b16}, the constitutive equation for $\mathbf{q}_i^{(1)}$ is
\begin{equation}
\label{b17}
\mathbf{q}_i^{(1)}=-T^2D_i''\nabla x_1-L_{p,i}\nabla p-\lambda_{T,i}\nabla T.
\end{equation}
From dimensional analysis, $D_i\propto T^{-1/2}$, $L_{p,i}\propto T^{3/2}/p$, and $\lambda_{T,i}\propto T^{1/2}$. Taking into account these results, $\partial_t^{(0)} \mathbf{q}_i^{(1)}$ can be explicitly written in terms of the spatial gradients of the fields as
\beq
\label{b18}
\partial_t^{(0)}{\bf q}_i^{(1)}=\left[\frac{3}{2}\zeta^{(0)}T^2D_i''+ \left( \frac{\partial \zeta ^{(0)}}{
\partial x_{1}}\right)_{p,T}(p L_{p,i}+T\lambda_{T,i})\right]\nabla x_1+\zeta^{(0)}\left(\frac{5}{2}L_{p,i}+\frac{T\lambda_{T,i}}{p}\right)\nabla p+
\zeta^{(0)}\left(\lambda_{T,i}-\frac{p L_{p,i}}{2T}\right)\nabla T.
\eeq
Substitution of Eq.\ \eqref{b18} into Eq.\ \eqref{b16} and taking into the constitutive equation \eqref{3.28} for the mass flux, one obtains the matrix equations \eqref{3.46} and \eqref{3.51} for the coefficients $D_i$, $L_{p,i}$, and $\lambda_{T,i}$.

\end{widetext}


%

\end{document}